\RequirePackage{fix-cm}
\documentclass[twocolumn]{svjour3}          
\smartqed  
\usepackage{graphicx}
\usepackage{natbib}
\usepackage{amsmath}
\begin{document}

\title{Characterization of biophysical determinants of spatio-temporal calcium dynamics in astrocytes}


\author{Thais Appelt Peres Barti\^e         \and
        Leonel Teixeira Pinto }
\authorrunning {Thais  Barti\^e  \and
                       Leonel Pinto }

\institute{Thais Appelt Peres Barti\^e \at
              \email{thais.appelt@gmail.com}           
           \and
           Leonel Teixeira Pinto \at
              \email{leonel.t.pinto@ufsc.br}           
\and
Department of Chemical Engineering and Food Engineering, Federal University of Santa Catarina, Florian\'opolis, Santa Catarina, Brazil \\}

\date{Received: date / Accepted: date}

\maketitle

\begin{abstract}
Most of the functions performed by astrocytes in brain information processing are related to calcium waves. Experimental studies involving calcium waves present discrepant results, leading to gaps in the full understanding of the functions of these cells. The use of mathematical models help to understand the experimental results, identifying chemical mechanisms involved in calcium waves and the limits of experimental methods. The model is diffusion-based and uses receptors and channels as boundary conditions. The computer program developed was prepared to allow the study of complex geometries, with several astrocytes, each of them with several branches. The code structure allows easy adaptation to various experimental situations in which the model can be compared. The code was deposited in the ModelDB repository, and will be under number 266795 after publication. A sensitivity analysis showed the relative significance of the parameters and identifies the ideal range of values for each one. We showed that several sets of values can lead to the same calcium signaling dynamics. This encourages the questioning of parameters to model calcium signaling in astrocytes that are commonly used in the literature, and it suggests better experimental planning. In the final part of the work, the effects produced by the endoplasmic reticulum when located close to the extremities of the branches were evaluated. We conclude that when they are located close to the region of the glutamatergic stimulus, they favor local calcium dynamics. By contrast, when they are located at points away from the stimulated region, they accelerate the global spread of signaling.

\keywords{Glutamatergic stimulus \and Ion channels dynamics \and Mathematical modeling \and Calcium signaling \and Astrocyte dynamics}
\end{abstract}

\section{Introduction}
\label{intro}
There is experimental evidence to suggest that neurons and astrocytes communicate with one another. This suggests that astrocytes are involved in brain information processing. Researchers have proposed several experimental approaches in order to understand the role of astrocytes in brain communication \citep{halassa2007tripartite, perea2009tripartite, pirttimaki2017astrocyte}. These approaches are primarily designed to analyze calcium signaling, because this is the main activity of astrocytes \citep{bernardinelli2014astrocyte, bindocci2017three, howarth2014contribution, navarrete2012astrocyte, sasaki2014astrocyte}. The results suggested some information regarding the role of astrocytes; however, they are quite discrepant. The discrepancies of the experimental results hinder development of a global view of the contributions of astrocytes in brain communication \citep{bazargani2016astrocyte, lallouette2018astrocyte, volterra2014astrocyte}. Recently, the very existence of gliotransmission was discussed in two perspectives papers \citep{fiacco2018DualPerspectives, savtchouk2018DualPerspectives}, with opposite conclusions. The former do not believe in its existence under physiological conditions, while the latter consider it to be well established. These discrepancies arise from difficulties in detecting and interpreting calcium signaling. Elaboration of sophisticated experimental apparatus and critical analysis of results can improve understanding of the role of astrocytes \citep{losi2017new, poskanzer2018dynamism, rusakov2015disentangling}. Mathematical models can contribute to experimental data interpretation, because they help to identify the mechanisms involved in specific phenomena as well as the limits of experimental methods. Furthermore, validated mathematical models allow the elaboration of new hypotheses about each phenomenon involved and the prediction of a system’s dynamic behavior in various situations \citep{min2012computational, myung2002mathematical, suffczynski2006some}.

Both mathematical descriptions and interpretation of experimental results of calcium signaling are not trivial. They face problems such as geometric complexity of astrocytes as well the large number of mechanisms involved in calcium signaling generation and propagation. Mathematical models also face another challenge: how to mathematically describe cellular processes such as the dynamics of the various channel types and their opening probabilities. The main problems associated with interpretation of calcium signaling and its mathematical description will be elaborated below.

The astrocytes are considered in this work. They are called "stars" due to their high level of branching \citep{matyash2010heterogeneity}. The branches are important because they allow astrocytes to contact neurons and other astrocytes. In addition, the branches allow the astrocytes to involve distant synapses, building communication paths between them \citep{halassa2007synaptic}. Lack of reflection on this structural complexity may result in antagonistic interpretations of astrocyte phenomena \citep{bazargani2016astrocyte, volterra2014astrocyte}. In order to minimize this antagonism, it is important to consider where the signaling is detected, because cell processes occur at different amplitudes and periods depending on the region where they start \citep{shtrahman2017understanding, wu2014spatiotemporal}. Another important question is the way astrocytic connections are spatially distributed. The spatial distribution can directly affect calcium signaling dynamic \citep{perez2014structural}.

Calcium signaling is triggered by a combination of various mechanisms \citep{mccarthy1991pharmacologically}. The endoplasmic reticulum (ER) is the main organelle contributing to signaling through calcium release by channels \citep{berridge2003calcium, bootman2002calcium}. In addition to IP\textsubscript{3}-activated Ca\textsuperscript{2+} channels, the endoplasmic reticulum has ryanodine-sensitive receptors. Both IP\textsubscript{3} and ryanodine receptors are modulated by a mechanism called calcium-induced calcium-release (CICR) as a consequence of the increase of intracellular Ca\textsuperscript{2+} concentrations \citep{simpson1998characterization,skupin2008does}. Other biological processes participate in calcium signaling in astrocytes, including channel opening triggered by activation of glutamate ionotropic receptors \citep{koh1995ca2+} and ATP \citep{neary1988atp}; activation of Na\textsuperscript{+}/Ca\textsuperscript{2+}exchanger \citep{golovina2003na+}, and Ca\textsuperscript{2+}-ATPase pump (PMCA) \citep{verkhratsky2012calcium}, and opening of voltage-dependent calcium channels \citep{macvicar1984voltage}. In addition, passive channels are present in cytoplasmic and ER membranes \citep{clapham1995calcium}. Also, there are channels activated by lack of Ca\textsuperscript{2+} in ER linked to transient receptor potential channels (TRPC) \citep{verkhratsky2014store}.
	
Based on several mechanisms described above, it is possible to conclude that calcium signaling is a complex phenomenon. Thus, development of a mathematical model that describes it requires selection of the most relevant mechanisms. This is not an easy task; however, good selection is fundamental such that the mathematical model may permit plausible inferences.

Several mathematical models of calcium signaling in astrocytes have been proposed in recent years; however, as was the case with experimental results, these models have not yet led to definitive conclusions regarding the role of astrocytes in brain information processing. Most of these models were derived from one of three classic models describing calcium signaling: \citet{keizer1992two}, \citet{li1994equations} or \citet{hofer2002control}. In addition to using one of these models as a basis, recent models used the same values as their parameters. \citet{keizer1992two}, for example, stated that the set of parameters they used was not the only one to satisfy the established conditions. In addition, parametric sensitivity analysis has been rarely performed in papers proposing mathematical modeling of calcium signaling in astrocytes \citep{manninen2018computational}. Lack of parametric analysis may be responsible for comparability problems observed by \citet{manninen2017reproducibility}. Another relevant question is why the three basis papers do not make direct comparisons with experimental results. Recent papers rarely do this either. It is possible that this phenomenon is related to the many gaps remain in experimental results \citep{poskanzer2018dynamism, durkee2018diversity, oheim2017local, shigetomi2016probing, rusakov2015disentangling}. It is also important to note that \citet{manninen2018computational} and \citet{manninen2017reproducibility} point to difficulties in reproducing literature models, either on account of errors in model creation itself or by absence of information.

Given the discrepancies and inconsistencies of experimental results and mathematical models, deep reflection and engagement in the creation of new approaches are necessary. These approaches must consider peculiarities of astrocytes and the various challenges they present.

Against this background, the objective of this work was to propose a biologically coherent mathematical model for prediction of calcium signaling dynamics triggered by glutamatergic stimuli in astrocytes. This model considers the most relevant characteristics and processes regarding astrocytes to be used as a tool for further studies on the roles of these cells. A parametric sensitivity analysis was performed and the source code of computational program will be available in ModelDB \citep{mcdougal2017twenty} at http://modeldb.yale.edu/266795, allowing for its reproduction. The code is available on the second author's ResearchGate. Program availability and the manner in which it was built enables researchers worldwide to contribute in many ways, both experimental and mathematical. As the model is being enriched, it will be possible to make effective contributions to understanding of the role of astrocytes in brain information processing.

\section{Mathematical model}
\label{math_model}

Calcium signaling depends on mechanisms that occur mainly in three regions: synaptic cleft, cytosol, and endoplasmic reticulum. In the first region, only the temporal dynamics of glutamate that reaches the astrocytes proximity is modeled. In the cytosol, the dynamics of calcium and IP\textsubscript{3} are modeled. The region is considered heterogeneous and the model includes diffusive partial differential equations. The endoplasmic reticulum acts as a functional calcium storage site. Due to the proportionally small dimensions of the ER and its high calcium concentrations, it is possible to disregard its concentration profiles and to use a homogeneous approach with ordinary differential equations. Homogeneity is used here, not in the mathematical sense, but in the chemical sense, as a mixture that has a uniform concentration at all points. If this consideration is possible, the benefit would be reduced computational time and memory requirements. The two approaches, homogeneous and heterogeneous, will be tested and compared.

\begin{figure*}
  \includegraphics[width=1.0\textwidth]{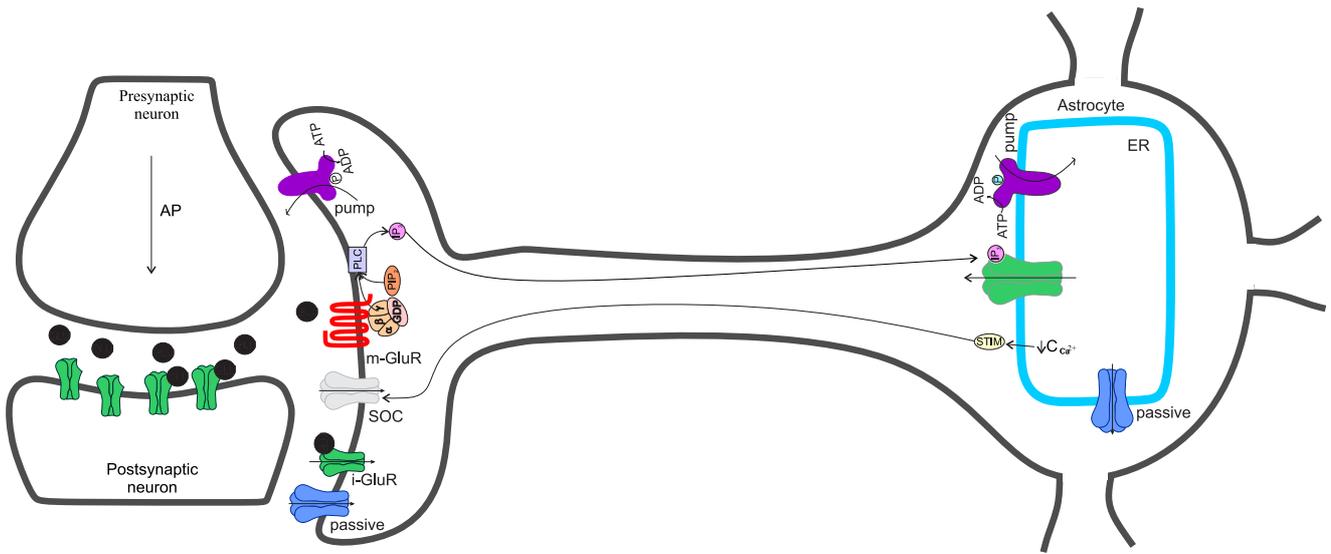}
\caption{\textbf{Calcium signaling mechanisms considered in the mathematical model.} The diagram illustrates some mechanisms involved in calcium signaling. In the cell membrane, there are metabotropic glutamate receptors (m-GluR), which are activated when neurons release glutamate into the synaptic cleft during synaptic transmission. Its activation triggers production of IP\textsubscript{3}, which moves by diffusion. Upon reaching the endoplasmic reticulum membranes, IP\textsubscript{3} activates receptors that open calcium channels. Calcium concentration decreases in the endoplasmic reticulum promote activation of STIM, which opens store-operated calcium channels (SOC) located in the cell membrane. In both cell and endoplasmic reticulum membranes, passive channels and pump operate to maintain or restore resting state. In the cell membrane, glutamate released by neurons also activates ionotropic receptors (i-GluR), opening calcium channels. All mechanisms mentioned above were included in the mathematical model.}
\label{fig:1}       
\end{figure*}

Fig. \ref{fig:1} shows a schematic drawing of regions included in the mathematical model, highlighting the mechanisms considered in each. As the objective of this work was to study calcium signaling triggered by glutamatergic stimulus, the main mechanisms considered were: activation of ionotropic and metabotropic glutamate receptors, i-GluR and m-GluR \citep{verkhratsky2012calcium}; activity of ATPases and Ca\textsuperscript{2+} passive channels in both cell and ER membranes \citep{clapham1995calcium}; opening of IP\textsubscript{3}-activated channels inthe ER membrane \citep{verkhratsky2012calcium}; TRPC channel operation, and a contributor to mechanism called store-operated calcium entry (SOCE) \citep{verkhratsky2014store}.

After choosing the relevant mechanisms, mass balance was performed in regions where ordinary or partial differential equations are used (cytosol and ER).

The cytosol is a colloidal medium that contains organelles and other structures that make the free movement of ions and molecules difficult \citep{luby1999cytoarchitecture}. In this medium, diffusion is the main mass transport mechanism \citep{hofer2002control}. Consequently, calcium and IP\textsubscript{3} dynamics in cytosol (\(C\) and \(I\), respectively), can be described by Eq. \ref{Equation 1}, where \(D_{W}\) is the compound \(W\) diffusion coefficient in cytosol (cyt):

\begin{equation}
\label {Equation 1}
\frac{\partial W_{cyt}}{\partial t} = D_W.\nabla^2{W_{cyt}}, W = C, I
\end{equation}

The same equation can be used to describe calcium dynamics in ER, when this organelle is considered heterogeneous. In the equations throughout the text, the variables are the concentrations of IP\textsubscript{3} (I), calcium (C) and glutamate (G). The main subscript used refer to the regions where the variables are calculated, cytosol (cyt), endoplasmic reticulum (ER) and synaptic cleft (sc) and to the membranes where flows are calculated, endoplasmic reticulum (ERM) and cell body (CM). The description of the parameters is presented in Table 1.

As previously mentioned, the endoplasmic reticulum can be considered homogeneous. Because of its narrow thickness, the diffusive flow of calcium quickly smooths out the concentration gradients created by outflow. In this case, the calcium fluxes occurring in its membrane must be directly computed in mass balance. Assuming the variation of calcium concentration in the ER depends on the fluxes produced by calcium pumps (\(v^{pp}_{ERM}\)) IP\textsubscript{3}-activated calcium channels (\(v^{lc}_{ERM}\)) and calcium passive channels (\(v^{pc}_{ERM}\)) \citep{ullah2006anti}, Eq. \ref{Equation 2} can be written to describe the calcium concentration profile in ER. It is necessary to calculate the integral of calcium flux over the surface of ER membrane (\(S_{ERM}\)) to reconcile the homogeneous ER approach with the heterogeneous cytosol approach.

\begin{equation}
\label {Equation 2}
V_{ER}\frac{\partial C_{ER}}{\partial t} = \sum_{g=1}^{3} \int\int_{S_{ERM}} v^{g}_{ERM}
\end{equation}
where g = 1(pp), 2(pc), 3(lc)\\

The consideration of the homogeneity of the mixture within the endoplasmic reticulum is typical in the mathematical models found in the literature. The review made by \citep{manninen2018computational} listed 106 mathematical models generated between 1995 and 2018. Only three of these consider diffusion of calcium in the ER. The question is not whether the flux of calcium through ER membranes creates concentration profiles, which obviously happens, but whether they need to be considered in the mathematical model. To help understand this issue, both approaches were analyzed. The same program configuration was performed for a duration of two seconds, with both considerations and their results were compared. The differences were less than 0.0001\%, considered insignificant. The differences in the concentration curves were not distinguishable, which is why they are not shown in this work. According to Occam razor, it was determined to maintain the consideration of homogeneity.

The buffering effects of calcium can be considered in two ways: by considering the free calcium concentration, or by considering the calcium effective diffusion coefficient in the cytosol. The effective diffusion coefficient was the form chosen for this work \citep{hofer2002control}.

Mathematical expressions for almost all of fluxes produced by the mechanisms represented in Fig. \ref{fig:1} can be found in literature. However, they were developed to be used in homogeneous models. Even in some already proposed heterogeneous models, such as thosw of \citet{hofer2002control} and \citet{wu2014spatiotemporal}, these fluxes were incorporated in mass balance. This was done to simplify the mathematical model and to avoid the need to incorporate these fluxes as boundary conditions. In this paper, we consider the places where fluxs occur, which implies their use as boundary conditions. They are represented by equations 3 to 5 and 7 to 9. Therefore, the mathematical expressions proposed in the literature have been modified to include the density of the channels and, therefore, the fraction of the area occupied by the channels in the membranes of the cell and the ER.

IP\textsubscript{3} production depends not only on glutamate coupling with metabotropic receptor, but also on the concentration of intracellular calcium, because Ca\textsuperscript{2+} keeps the effector protein activated \citep{pawelczyk1997structural}. \citet{ullah2006anti} shows an example for calculation of IP\textsubscript{3} production summing the effects of these two compounds. This appears inappropriate, given that the protein amplifies, but does not initiate the activation \citep{thore2005feedback}. Based on this fact, the present model multiplied the two contributions, as shown in Eq. \ref{Equation 3}. In relation to the term corresponding to calcium contribution, the proposal of \citet{hofer2002control} was used. In the equation the gradient \((G_{sc}-G^*_{sc})\) is used because it is assumed that IP\textsubscript{3} production is not relevant at rest concentration. Thus, Eq. \ref{Equation 3} represents the IP\textsubscript{3} production in an infinitesimal element of area, dA.

\begin{equation}
\label {Equation 3}
u^{p}_{CM}=f^{p}_{CM}.dA_{CM}K^p\frac{(G_{sc}-G^*_{sc})^n}{(G_{sc}-G^*_{sc})^n+k_{p,G}^n}\frac{C_{cyt}^2}{C_{cyt}^2+k_{p,C}^2}
\end{equation}

IP\textsubscript{3} degradation, Eq. \ref{Equation 4}, was described using first-order kinetics, as proposed by \citet{ullah2006anti}. Elimination of IP\textsubscript{3} occurs until its concentration returns to the resting value. 

\begin{equation}
\label {Equation 4}
u^{d}_{ERM}=f^{d}_{ERM}.dA_{ERM}.h^d.(I_{cyt}-I^*_{cyt})
\end{equation}

Calcium flux through passive channels, both in the cellular or ER membranes \citep{di2007calcium}, was considered proportional to the area fraction of these channels, to mass transference coefficient, and to concentration gradient between both sides of the membranes. This relation replaces the concept of constant current of calcium entry in the cell, which is used in several articles \citep{amiri2011functional, hofer2002control, ullah2006anti}. Eq. \ref{Equation 5} shows calcium flux through passive channels in the ER membrane.

\begin{equation} \label {Equation 5}
v^{pc}_{ERM}=f^{pc}_{ERM}.dA_{ERM}.h^{pc}_{ERM}.(C_{ER}-C_{cyt})
\end{equation}

We incorporated transient receptor potential channels (TRP channels) activated from a reduction in ER calcium concentration. It is possible consider these channels present identical dynamics to those of passive channels after activation and opening. It can be assumed that there is an effective increase in passive channel fraction in the cell membrane that depends on Ca\textsuperscript{2+} concentration in the ER. Mathematically, this idea is expressed in the Eq. \ref{Equation 6}.

\begin{equation} \label {Equation 6}
f^{TRPC}_{CM}=\left\{\begin{array}{rc}
\frac{1}{100}.\frac{C^*_{ER}-C_{ER}}{C^*_{ER}},&\mbox{for}\quad {C^*_{ER}-C_{ER}}> 0,\\
0, &\mbox{for}\quad {C^*_{ER}-C_{ER}} \le 0.
\end{array}\right.
\end{equation}

The reduction factor (100) employed in Eq. \ref{Equation 6} causes the fraction of these channels to be at the same order of magnitude as those of other channels.

Eq. \ref{Equation 7} gives the adaptation of Eq. \ref{Equation 5} to the cytoplasmic membrane, taking into account the TRP channels.

\begin{equation} \label {Equation 7}
v^{pc}_{CM}=(f^{pc}_{CM}+f^{TRPC}_{CM}).dA_{CM}.h^{pc}_{CM}.(C_{sc}-C_{cyt})
\end{equation}

The equation for describing calcium flux produced by the pumps in the ER membrane was developed from experimental data fit. The equation considers that pump operation depends of intracellular calcium concentration, up to a saturation limit, given by the \(k_{pp}\) constant  \citep{ullah2006anti}. Calcium influence is of second order, because the pump transports two calcium ions simultaneously. According to \citet{verkhratsky2012calcium}, these pumps are also located in cell membrane. Its action was included in the present model. Eq. \ref{Equation 8} describes the calcium flux due to pump action.

\begin{equation}
\label {Equation 8}
v^{pp}_{m}=f^{pp}_{m}.dA_{m}.K^{pp}_{m}.\frac{C^2_{cyt}}{C^2_{cyt}+k^2_{pp}}, m = CM, ERM
\end{equation}

Eq. \ref{Equation 9}, based on \citep {ullah2006anti}, represents calcium flux through ligand-activated channels present in ER and cell membranes.

\begin{equation}
\label {Equation 9}
v^{lc}_{m}=p^{o/c}_m.f^{lc}_{m}.dA_{m}.h^{lc}_{m}(C_{r}-C_{cyt})\left\{\begin{array}{rc}
m = CM, r = sc   \\
m = ERM, r= ER
\end{array}\right.
\end{equation}

The mathematical description of opening and closing of ionic channels, among all the previously described phenomena, is the most challenging, because the channels are not simultaneously activated \citep{falcke2003role} and each one has its own particularities \citep{eisenberg1999structure}. There are interesting proposals in several papers; nevertheless, problems remain, including the number of parameters and the absence of clear criteria for the choices of their values \citep{keizer1992two, means2006reaction}. Although they have been used in later investigations \citep{hofer2002control, taheri2017diversity, ullah2006anti, wu2014spatiotemporal}, these values were obtained in studies of different cell types; therefore, their generalization for study of astrocytes is questionable.

\begin{figure*}
  \includegraphics[width=1.0\textwidth]{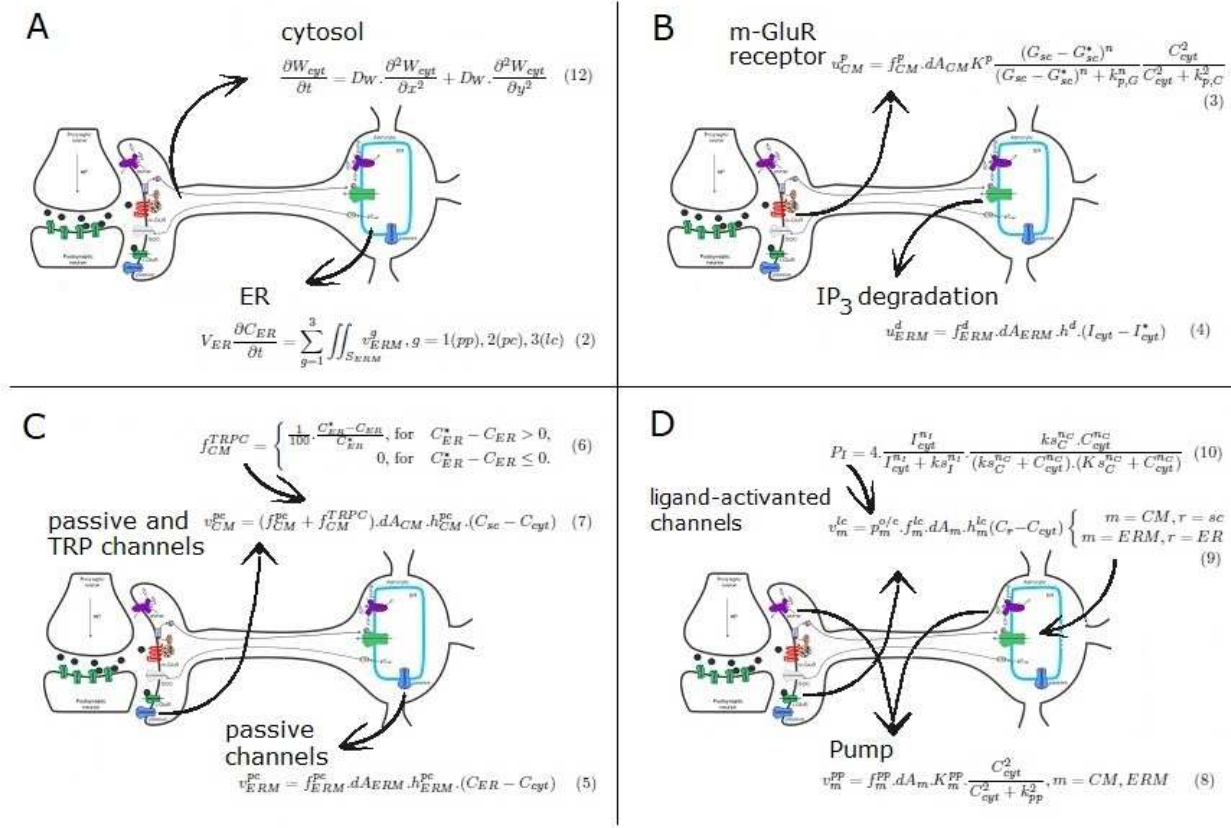}
\caption{\textbf{equations application domain} The diagram illustrates some mechanisms involved in calcium signaling. A. Eq. \ref{Equation 12} describes the diffusive movement of Ca\textsuperscript{2+} and IP\textsubscript{3} in the cytosol of all regions. It is known as Fick's second law of diffusion. Eq. \ref{Equation 2} is used to calculate the variation in ER Ca\textsuperscript{2+} concentration. It says that any variation in this concentration is result of Ca\textsuperscript{2+} fluxes that occur in the ER membrane, whether produced by calcium pumps, passive channels or ligand-activated channels. B. IP\textsubscript{3} is formed by m-GluR receptors, as a result of glutamatergic stimulus (Eq. \ref{Equation 3}). It moves through the cytosol by diffusion (Eq. \ref{Equation 12}) and is degraded in the ER membrane (Eq. \ref{Equation 4}). C. Calcium flux through passive channels is calculated on ER membranes (Eq. \ref{Equation 5}) and on plasmatic membranes at the ends of the branches (Eq. \ref{Equation 7}). The difference is that the latter includes Transient Receiver Potential Channels, TRPC. The lateral membranes of the branches are considered isolated. D. The fluxes produced by calcium pumps (Eq. \ref{Equation 8}) and ligand-activated channels (Eq. \ref{Equation 9}) are calculated for both membranes. The probability of opening the channels is calculated by Eq. \ref{Equation 10}.}
\label{fig:2}       
\end{figure*}

The discussion about opening and closing of ionic channels is related to the dynamics of the parameter \(p^{o/c}_{m}\) in Eq. \ref{Equation 9}, indicating whether the channel allows, at a specific moment, the passage of calcium (\(p^{o/c}_{m}=1\)) or not  (\(p^{o/c}_{m}=0\)). Results obtained by \citet{tu2005modulation} and \cite{tu2005functional} were used for description of dynamics parameters. These researchers conducted a detailed study of IP\textsubscript{3} activated channels and they proposed equations that describe the channel probability of opening according to cytosolic concentrations of Ca\textsuperscript{2+} or IP\textsubscript{3}, respectively \(P_{ca}\) and \(P_{I}\). The authors calculated the probability of opening either maintaining constant IP\textsubscript{3} concentration for the first, or Ca\textsuperscript{2+} concentration for the last. 

They obtained two experimental curves, fitted with specific functions. In this model, both concentrations vary simultaneously. The total probability of opening channels must meet the condition \(P_{I} \in [0;1]\). Applying the theorem for independent probabilities, \(P(A\cap B)=P(A)\) x \(P(B)\), the two expressions proposed by Tu et al. have been multiplied. The result is Eq. \ref{Equation 10}, which represents the total probability of opening channels activated by IP\textsubscript{3}, for the situation of simultaneous and independent variation of calcium and IP\textsubscript{3} concentrations. 

\begin{equation}
\label {Equation 10}
P_{I}=4.\frac{I^{n_I}_{cyt}}{I^{n_I}_{cyt}+ks^{n_I}_{I}}.\frac{ks^{n_C}_{C}.C^{n_C}_{cyt}}{(ks^{n_C}_{C}+C^{n_C}_{cyt}).(Ks^{n_C}_{C}+C^{n_C}_{cyt})}
\end{equation}

For glutamate-activated calcium channels (i-GluR), a similar equation was used. Because there is no information on the participation of Ca\textsuperscript{2+} in the activation or inactivation process, a simpler equation was used, solely dependent on glutamate. The Eq. \ref{Equation 11} is similar to the first term of \(P_{I}\); however, the concentration of glutamate is now considered.

\begin{equation}
\label {Equation 11}
P_{G}=\frac{G^{n_G}_{sc}}{G^{n_G}_{sc}+ks^{n_G}_{G}}
\end{equation}

Fig. \ref{fig:2} shows in detail the locations where each equation in the mathematical model is used. It mainly seeks to detail the application of boundary conditions.

\section{Aspects related to computer implementation}
\label{sec:2}

The geometry to represent the astrocytes should basically contain several branches and a body to which the branches can be coupled. Shapes, such as prisms and spheres, could be used to represent cell body geometry, facilitating the use of macroscopic equations to describe molecular motion. Prisms or cylinders can represent the branches. The computational implementation of the mathematical model considered that most of available data comes from experiments with astrocytes cultures seeded in monolayers. The use of a two-dimensional model \citep{hofer2002control} is reasonable in these conditions, what makes possible to consider the cell body and its branches as rectangular prisms containing the same depth. The Cartesian coordinate system is appropriate to the chosen geometry. Thus, considering this coordinate system and two-dimensional variation,  Eq. \ref{Equation 1} can be rewritten as the Eq. \ref{Equation 12}.

\begin{equation}
\label {Equation 12}
\frac{\partial W_{cyt}}{\partial t} = D_W.\frac{\partial^2 W_{cyt}}{\partial x^2}+D_W.\frac{\partial^2 W_{cyt}}{\partial y^2}, W = C, I
\end{equation}

The location and geometry of endoplasmic reticulum influence calcium signaling \citep {berridge2003calcium}. According to \citet {pivneva2008store}, the endoplasmic reticulum locates in vicinity of the nucleus and is also dispersed in the cytosol, near the plasma membrane. In astrocytes in situ, the investigators have found most of reticulum in vicinity of plasma membrane, that is, in the branches. \citet{patrushev2013subcellular}, By contrast, points out that the volume occupied by ER increases from the end of branches towards the cell body. In addition, \cite{oschemann2016cellnetwork}  demonstrated that endfeet processes are triggered mainly by calcium input from the extracellular environment and not by release of calcium stored in RE. Literature shows that there is no single opinion on this subject. 

Based on this information this work was carried out in two steps. The first stage considered the study of effects produced by cytosol scattered ERs requiring a great computational effort. Not only by increasing the number of variables, but also by the number of possible alternatives, because the positions of ERs become almost random. 

\begin{figure}
  \includegraphics[width=0.48\textwidth]{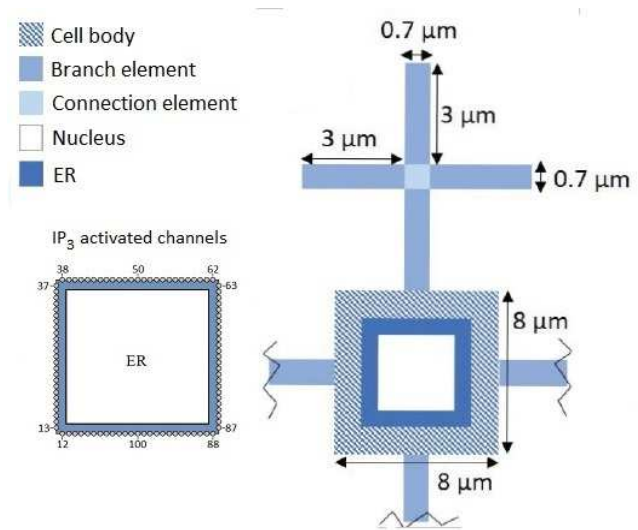}
\caption{\textbf{Cell geometry used in mathematical modeling.} The diagram shows the structure used. It consists of four branches connected to the cell body. The cell body has a nucleus surrounded by the endoplasmic reticulum. The dimensions, in \(\mu m\), are, 8.0 x 8.0 x 1.0 for the cell body, 3.0 x 0.7 x 1.0 for branch elements, 0.7 x 0.7 x 1.0 for connecting elements, and 4.3 x 4.3 x 1.0 for the nucleus. The nucleus is surrounded by the ER, whose external faces, with dimensions of 5.4 x 1.0, have one hundred of IP\textsubscript{3}-activated channels. These channels are numbered clockwise from the center of the bottom face. The ER is isolated on its internal faces.}
\label{fig:3}       
\end{figure}

Given this, were performed calcium signaling analyzes without the existence of ER in the branches. Each cell has a single ER, positioned around the nucleus. Nucleus and ER shapes and positions are shown in Fig. \ref{fig:3}. In the second stage a simplified structure was elaborated, in which astrocytes are composed by a central region with two opposite arms. The ER positioned in the branches are formed by rectangular prisms. This configuration will be detailed later.

Fig. \ref{fig:3} shows the representation of the computationally implemented geometry, highlighting the dimensions and the distribution of the branches.

It is important to note that the computer program developed is versatile enough to build several cells, each with several complex branches, of different sizes and in different positions. This versatility is best described in a file attached to the software in ModelDB.

For the solution of Eq. \ref{Equation 12}, a mesh was constructed filling all the cytosol area, with \(\Delta x\) and \(\Delta y\) dimensions. In the computer program, \(\Delta x=0.1 \mu m\) and \(\Delta y=0.05 \mu m\) were chosen in all regions. The fluxes produced by the channels illustrated in Fig. \ref{fig:1} only reach this mesh in its contours. For this reason, they must be considered as boundary conditions. These conditions will now be detailed

\subsection{Boundary conditions}
\label{sec:3}

The boundary conditions describe the effect of calcium and IP\textsubscript{3} fluxes on the control volumes that belong to cell membranes. As shown in Fig. \ref{fig:1}, these membranes delimit the branches, the cell body and the ER. The branches lateral faces are considered insulated. At its extremities, by contrast, there are calcium fluxes through channels activated by glutamate, passive channels and pump. In addition, there is production flux of IP\textsubscript{3}. The boundary condition in these cases is based on the idea that mass flux crossing membrane moves by diffusion within the cell. Mathematically, this condition is represented by Eq. \ref{Equation 13} (IP\textsubscript{3}) and Eq. \ref{Equation 14} (Ca\textsuperscript{2+}).

\begin{equation}
\label {Equation 13}
v^{p}_{CM} = -dA_{CM}.D_I.\frac{\partial I_{cyt}}{\partial u}
\end{equation}

\begin{equation}
\label {Equation 14}
v^{lc}_{CM}+v^{pc}_{CM}+v^{pp}_{CM} = -dA_{CM}.D_C.\frac{\partial C_{cyt}}{\partial u}
\end{equation}

In membrane surrounding the cell body, the existing fluxes are calcium through passive channels and the PMCA pump. Considering again the movement inside the cell occurring by diffusion, the boundary condition at this location is given by Eq. \ref{Equation 15}.

\begin{equation}
\label {Equation 15}
v^{pc}_{CM}+v^{pp}_{CM} = -dA_{CM}.D_C.\frac{\partial C_{cyt}}{\partial u}
\end{equation}

Finally, in membrane surrounding the endoplasmic reticulum there are calcium fluxes through channels activated by IP\textsubscript{3}, and through passive channels and pumps. In addition, there is degradation flux of IP\textsubscript{3}. The boundary conditions at this location, are given by Eq. \ref{Equation 16} (IP\textsubscript{3}) and Eq. \ref{Equation 17} (Ca\textsuperscript{2+}), considering again diffusion in cytosol.

\begin{equation}
\label {Equation 16}
v^{d}_{ERM} = -dA_{ERM}.D_I.\frac{\partial I_{cyt}}{\partial u}
\end{equation}

\begin{equation}
\label {Equation 17}
v^{lc}_{ERM}+v^{pc}_{ERM}+v^{pp}_{ERM} = -dA_{ERM}.D_C.\frac{\partial C_{cyt}}{\partial u}
\end{equation}

The two sets of equations in the model, equations 3-9 and equations 13-17 are complementary. We will show its application in the model using equations 4 and 16, which represent the degradation of IP\textsubscript{3} in the ER membrane. Substituting the first in the second and simplifying, we obtain:

\begin{equation}
\label {Equation 18}
\frac{\partial I_{cyt}}{\partial u}=-\frac{f^{d}_{ERM}.h^{d}}{D_{I}}.(I_{cyt}-I^*_{cyt}), u=x,y
\end{equation}

Eq. \ref{Equation 18} means that any IP\textsubscript{3} that arrives by diffusion in the ER membrane, activates calcium channels and is degraded. The discretization of the equations was performed using the Finite Difference Method. Considering a point \((i,j)\) belonging to the mesh used in the cytosol to discretize Eq. \ref{Equation 12}, and assuming that point \(i\) belongs to the ER membrane, and that a regressive discretization is used at this point, we would have:

\begin{equation}
\label {Equation 19}
I_{i-1,j}=I_{i,j} + \alpha.(I_{i,j}-I^*_{cyt})
\end{equation}

where the dimensionless \(\alpha = f^{d}_{ERM}.h^{d}.\Delta x/D_{I}\)

Note that, in the discretization of Eq. \ref{Equation 12}, applied at point \(i\) belonging to the ER membrane, the concentration of I appears in position \((i-1,j)\). This position does not belong to the equation domain, which is why it needs to be replaced by the boundary condition.


\subsection{Opening probability}
\label{sec:4}

The probability of channel opening requires an additional explanation about how it was code implemented. This is necessary because probability value calculation is not enough to determine whether the channel will be opened or not. Another criterion must be established in order to decide the event occurrence. It was decide the channel would be opened only if a number,\(\gamma \in\) (0;1), randomly chosen from an equiprobable distribution, was less than the probability calculated for each computational step by Eq. \ref{Equation 10} or Eq. \ref{Equation 11}. The channel, once opened, remains on this condition for a period of time,  \(t_{o}\). After that, it is closed for another period, corresponding to the refractory period,  \(t_{r}\). Under these specific conditions there is no further need to calculate the probability, because \(p^{o/c}_{m}\) is equal to one or zero, respectively.

However, the probability of opening is dependent on the time interval used for each computational step. This is because, if two time intervals were used, the first being double the second, in the second the number of tests doubles and the probability of opening also doubles. To avoid this, all simulations were carried out with the same time interval, \(\Delta t = 10^{-6}s\)

\subsection{Model parameters values}
\label{sec:5}

It was assumed that the rest concentrations of calcium and IP\textsubscript{3} in cytosol and calcium in the ER correspond to 0.1 \(\mu\)M, 0.16 \(\mu\)M and 10.3 \(\mu\)M, respectively \citep{ullah2006anti}. In addition, calcium concentration in the extracellular fluid corresponds to 2 mM \citep{bear2007neuroscience}. Calcium diffusivity is 25 \(\mu m^2.s^{-1}\) and the IP\textsubscript{3} diffusivity, 280 \(\mu m^2.s^{-1}\) \citep{hofer2002control}.

The volumetric fraction occupied by reticulum was estimated based on data provided by \citet{alberts2017molecular}. The authors declared that cytosol occupies 54\% of cell total volume. Thus, all organelles occupy 46\% of cell body. As the cell fraction occupied by ER is about 17\%, the organelles fraction corresponding to ER is 37\%.

The calculation of the area fraction occupied by metabotropic receptors was based on the estimation of receptors density presented by \citet{nadkarni2005synaptic}: 300 \(\mu m^{-2}\). According to this information and because this type of receptor has one pore of approximately 5 nm of diameter \citep{traynelis2010glutamate}, the \(f^{p}_{CM}\) parameter was calculated (0.006). The same value was used to  \(f^{lc}_{CM}\),  \(f^{lc}_{ERM}\) and  \(f^{d}_{ERM}\) for lack of better information. The  \(k_{p,C}\) value (0.3 \(\mu\)M) was extracted from \citet{hofer2002control} and \(n\) (0.3) and \(k_{p,G}\) (0.78 \(\mu\)M) were from \citet{ullah2006anti}.

The constant associated to ER passive channels, \(f^{pc}_{ERM}\).\(h^{pc}_{ERM}\), was estimated by equaling the flux through channels used by \citet{ullah2006anti} to the rest flux used in this work. The obtained value was 0.12 \(\mu M.s^{-1}\). Same value was used for cytoplasmic membrane channels, i.e.\(f^{pc}_{CM}\).\(h^{pc}_{CM}\)=0.12 \(\mu M.s^{-1}\). This flux is ten times lower for cell body. Otherwise, the exit would be massive and the restitution period for ER would be too long, which is not biologically feasible.

The pump saturation constant, \(k^{pp}\) = 0.1 \(\mu\)M, was also used by \citet{ullah2006anti}. The constants \(f^{pp}_{ERM}\). \(K^{pp}_{ERM}\) and \(f^{pp}_{CM}\).\(K^{pp}_{CM}\) were calculated by equaling the flux of passive channels and the flux of pumps at resting. Similarly, to passive channels, this flux is ten thousand times lower in cell body membrane.

The opening probability constants for IP\textsubscript{3}-activated channels were extracted from  \citet{tu2005modulation} and \citet{tu2005functional}, considering that the astrocytes present, mainly type 2 IP\textsubscript{3} receptors \citep{holtzclaw2002astrocytes}. Thus, \(ks_I\) = 0.1 \(\mu M\); \(n_I = 2.2\); \(ks_C\) = 0.16 \(\mu M\); \(n_C = 2.05\) and \(Ks_C\) = 0.16 \(\mu M\). For glutamate-activated channels, initially, \(n_G\) equal to \(n_I\) was adopted. The value \(ks_G\) = 1000 \(\mu\)M was determined based on knowledge of interval in which glutamate concentration varies in the synaptic cleft. The concentration of glutamate at rest in the cleft is 3 \(\mu\)M. Given that, on average, 3000 glutamate molecules are released in cleft during synaptic transmission and cleft has a volume of 2.0 x \(10^6\) \(nm^3\) \citep{allam2012computational}, the average concentration of this compound reaches 3003 \(\mu\)M. Considering the glutamate release through the vesicle in a synapses central point, this concentration in the areas surrounding the astrocyte has an amplitude of \(\delta\)=1000 \(\mu\)M \citep{allam2012computational}. By plotting a curve with 1000, 2000, 3000 \(\mu\)M for \(ks_G\) value, the most coherent behavior was found at 1000 \(\mu\)M. Refractory period (10.8 ms) and opening maintenance time (7.6 ms) of each channel were extracted from \citet{tu2005functional}.

The constants associated with production and decay rates of IP\textsubscript{3}, \(K^{p}\) = 6.0 x \(10^{-12}\) \(\mu mol.\mu m^{2}/s\) and  \(h^{d}\) = 6.0 x \(10^{3}\) \(\mu m.s^{-1}\), were adjusted to make calcium signaling occurs in about 1.0 s as indicated in \citet{araque2014gliotransmitters}. The \(h^{lc}_{ERM}\) (150 \(\mu m/s\)) and \(h^{lc}_{CM}\) (0.3 \(\mu m/s\)) coefficients were adjusted assuming to be relevant the reticulum contribution to increase of intracellular calcium concentration and considering the comparative study with a calcium signaling presented by \citet{araque2014gliotransmitters}, this concentration doubles. It is noteworthy that after calculation of relation between \(v^{lc}_{CM}\) and \(v^{lc}_{ERM}\) flows obtained by \citet{ullah2006anti}, \(h^{lc}_{CM}\) corresponded to 0.002 x \(h^{lc}_{ERM}\).

Now, after established the global parameters, the results obtained with the model and its potentiality will be exposed.

\begin{table*}
\caption{Parameters description and adopted values}
\label{tab:1}       
\begin{tabular}{ p{1 cm} p{8 cm} p{2.6 cm}  p{1.6 cm}  p{2 cm} }
\hline\noalign{\smallskip}
Symbol & Description & Range & Value & Unit\\
\noalign{\smallskip}\hline\noalign{\smallskip}
\multicolumn{5}{c}{Rest concentrations}\\
\(C_{cyt}^{*}\)   & \raggedright{Calcium concentration at rest in cytosol} & & 0.1 & \(\mu M\) \\
\(C_{ER}^{*}\)   & \raggedright{Calcium concentration at rest in ER} & & 10.3 & \(\mu M\) \\
\(C_{sc}\)           & \raggedright{Calcium concentration in synaptic cleft} & & 2.0 & \(mM\) \\
\(I_{cyt}^{*}\)    & \raggedright{IP\textsubscript{3} concentration at rest in cytosol} & & 0.16 & \(\mu M\) \\
\(G_{sc}^{*}\)    & \raggedright{Glutamate concentration at rest in synaptic cleft} & & 3.0 & \(\mu M\) \\
\noalign{\smallskip}\hline\noalign{\smallskip}
\multicolumn{5}{c}{Diffusivities}\\
\(D_{C}\)            & \raggedright{Calcium diffusivity in cytosol} & & 25 & \(\mu m^2 s^{-1}\) \\
\(D_{I}\)             & \raggedright{IP\textsubscript{3} diffusivity in cytosol} & & 280 & \(\mu m^2 s^{-1}\) \\
\noalign{\smallskip}\hline\noalign{\smallskip}
\multicolumn{5}{c}{IP\textsubscript{3} production}\\
\(f^{p}_{CM}\)    & \raggedright{Cell membrane area fraction where occurs IP\textsubscript{3} production} & & 0.006 &- \\
\(K^{p}\)            & \raggedright{Maximum rate of IP\textsubscript{3} production} &  1 to \(20x10^{-12}\)  & \(6.0x10^{-12}\) & \(\mu mol  \mu m^{-2}  s^{-1}\) \\
\(k_{p,C}\)         & \raggedright{Dissociation constant for calcium stimulation of IP\textsubscript{3} production} & & 0.3 & \(\mu M\) \\
\(k_{p,G}\)         & \raggedright{Dissociation constant for glutamate stimulation of IP\textsubscript{3} production} & & 0.78 & \(\mu M\) \\
\(n\)                  & \raggedright{Hill coefficient for IP\textsubscript{3} production}   & & 0.3 & - \\
\noalign{\smallskip}\hline\noalign{\smallskip}
\multicolumn{5}{c}{IP\textsubscript{3} degradation}\\
\(f^{d}_{ERM}\) & \raggedright{Area fraction of ER membrane where occurs IP\textsubscript{3} degradation} & & 0.006 & -\\
\(h^{d}\) & \raggedright{IP\textsubscript{3} degradation coefficient} &  1 to 20 x  \(10^{3}\) & 6.0 x \(10^{3}\) & \(\mu m  s^{-1}\) \\
\noalign{\smallskip}\hline\noalign{\smallskip}
\multicolumn{5}{c}{Calcium fluxes through ligand channels }\\
\(f^{lc}_{CM}\)  & \raggedright{Cell membrane area fraction occupied by ligand channels} & & 0.006 & -\\
\(f^{lc}_{ERM}\) & \raggedright{ER membrane area fraction occupied by ligand channels} & & 0.006 & -\\
\(h^{lc}_{CM}\)  & \raggedright{Mass transfer coefficient for cell membrane ligand channels}  & & 0.3 & \(\mu ms^{-1}\) \\
\(h^{lc}_{ERM}\) & \raggedright{Mass transfer coefficient for ER membrane ligand channels}  & 10 to 200 & 150 & \(\mu ms^{-1}\) \\
\noalign{\smallskip}\hline\noalign{\smallskip}
\multicolumn{5}{c}{Calcium fluxes through passive channels }\\
\(f^{pc}_{CM}\) & \raggedright{Cell membrane area fraction occupied by passive channels} & & 0.001 & -\\
\(f^{pc}_{ERM}\) & \raggedright{ER membrane area fraction occupied by passive channels} & & 0.001 & -\\
\(h^{pc}_{CM}\) & \raggedright{Mass transfer coefficient for cell passive channels} & & 120 & \(\mu ms^{-1}\) \\
\(h^{pc}_{ERM}\) &\raggedright{Mass transfer coefficient for ER passive channels}  & & 120 & \(\mu ms^{-1}\) \\
\noalign{\smallskip}\hline\noalign{\smallskip}
\multicolumn{5}{c}{Calcium fluxes through pumps}\\
\(f^{pp}_{CM}\) & \raggedright{Cell membrane area fraction occupied by calcium pumps} & & 0.001 &- \\
\(f^{pp}_{ERM}\) & \raggedright{ER membrane area fraction occupied by calcium pumps} & & 0.001 & -\\
\(K^{pp}_{CM}\) &  \raggedright{Pump flux maximum rate at cell membrane}
& & 0.3 & \(\mu mol \mu m^{-2}s^{-1}\) \\
\(K^{pp}_{ERM}\) & \raggedright{Pump flux maximum rate at ER membrane} & & 0.3 & \(\mu mol \mu m^{-2} s^{-1}\) \\
\(k_{pp}\) &  \raggedright{Pump saturation constant} & & 0.1 & \(\mu M\) \\
\noalign{\smallskip}\hline\noalign{\smallskip}
\multicolumn{5}{c}{Channel open probability}\\
\(ks_{C}\) & \raggedright{Apparent affinity for calcium inhibitory site}  & & 0.16 & \(\mu M\) \\
\(Ks_{C}\) & \raggedright{Apparent affinity for calcium activating site} & & 0.16 & \(\mu M\) \\
\(ks_{I}\) &\raggedright{Apparent affinity for  IP\textsubscript{3} activating site}  & & 0.1 & \(\mu M\) \\
\(ks_{G}\) &\raggedright{Apparent affinity for glutamate activating site}  & & 1000 & \(\mu M\) \\
\(n_{C}\) &  \raggedright{Hill coefficient to calcium} & & 2.05 & - \\
\(n_{I}\) & \raggedright{Hill coefficient to IP\textsubscript{3}} & & 2.2 &-  \\
\(n_{G}\) &\raggedright{Hill coefficient to glutamate}  & & 2.2 &-  \\
\noalign{\smallskip}\hline\noalign{\smallskip}
\multicolumn{5}{c}{Channel opening and closing}\\
\(t_{r}\) & Ionic channels refractory time & & 10.8 & \(ms\) \\
\(t_{o}\) & Ionic channels opening time & & 7.6 & \(ms\) \\
\(t^{p}_{test}\) & Computing time for test probability & & 10.0 & \(ms\) \\
\noalign{\smallskip}\hline\noalign{\smallskip}
\multicolumn{5}{c}{Stimulus characteristics}\\
\(\sigma\) & \raggedright{Glutamate oscillation amplitude in synaptic cleft during a stimulus} & & 10.0 & \(ms\) \\
\(\delta\) & \raggedright{Glutamate oscillation amplitude during a stimulus} & & 1000 & \(\mu M\) \\
\noalign{\smallskip}\hline
\end{tabular}
\end{table*}

\section{Results and discussion}
\label{sec:6}

The calcium and IP\textsubscript{3} dynamics in the cytosol, as well as calcium dynamic in the ER, were obtained from numerical simulation of the proposed model and they are shown in this section. For this simulation, it was developed a computational interface using FORTRAN\textsuperscript{\textregistered} language.
  
The free parameters of the model were adjusted aiming to reach medium values indicated by \citet{araque2014gliotransmitters} for calcium signaling amplitude and duration. It was considered an astrocyte branch extremity stimulated due to passage of a signaling from a synapse which it surrounds. Here, the stimulus was considered glutamatergic, making necessary to consider glutamate concentration in synaptic cleft near the astrocyte. According to \cite{budisantoso2013synapsecleft} the concentration of Glutamate in the synaptic cleft grows exponentially close to the opening of a vesicle. The article also shows that, due to the spread of the neurotransmitter caused by diffusion, this growth softens as we move away from this point. Next to an astrocyte connected to the synapse, at a distance greater than 100 nm from the starting point, the growth of the glutamate concentration curve approaches a normal curve. The decay of the curve depends on diffusion and on the density of m-GluR and i-GluR channels. Given the difficulty of obtaining data on the dynamic behavior of glutamate close to the astrocytic termination, it was proposed a sine wave function shown in Eq. \ref{Equation 20}. This function has a behavior similar to that suggested by \cite{budisantoso2013synapsecleft} close to the astrocyte and guarantees a stimulus with a duration of 10 ms (\(\sigma\)) and a concentration amplitude of the order of \(\delta\). Another advantage is to easily allow the generation of stimulus sequences, or a burst. The sine function represents an increase followed by a decrease in glutamate concentration. For negative values of this function, the synaptic cleft glutamate concentration assumes the rest concentration value.

\begin{equation}
\label {Equation 20}
G_{sc} = G^{*}_{sc} +\delta.sin(\frac{2\pi}{\sigma}.t), \sigma = 10 ms
\end{equation}

For the graphs shown in this section, it was used the following dimensionless variables:

\begin{equation}
\label {Equation 21}
\overline{y}^{C}_{cyt}= \frac{\overline{C}_{cyt}}{C^{*}_{cyt}},   \overline{y}^{C}_{ER}= \frac{C_{ER}}{C^{*}_{ER}}, \overline{y}^I_{cyt}= \frac{\overline{I}_{cyt}}{I^{*}_{cyt}}
\end{equation}

\subsection{Cell calcium signalling}
\label{sec:7}

\begin{figure}
  \includegraphics[width=0.48\textwidth]{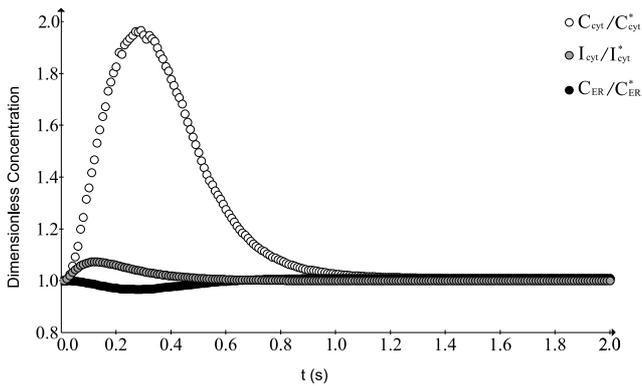}
\caption{\textbf{Calcium and IP\textsubscript{3} dynamics.} The graph shows calcium and IP\textsubscript{3} mean concentrations in cytosol and calcium concentration in endoplasmic reticulum. IP\textsubscript{3} concentration increases due to its production triggered by activation of metabotropic receptors in cell membrane. When it reaches endoplasmic reticulum, IP\textsubscript{3} starts to be degraded and its concentration decreases until it returns to resting value. Cytosol calcium concentration increases mainly because IP\textsubscript{3} reaches ER and opens calcium channels IP\textsubscript{3}-activated, allowing its release into cytosol. Calcium concentration in ER decreases because calcium is released into cytosol. Both calcium concentration in cell body and in ER  return to resting value due to calcium activity pumps and to passive channels.}
\label{fig:4}       
\end{figure}

The curves representing average concentration profiles of Ca\textsuperscript{2+} and IP\textsubscript{3} in the cytosol and Ca\textsuperscript{2+} in the ER (Fig. \ref{fig:4}) show coherent qualitative behavior. The glutamatergic stimulation activates metabotropic receptors, triggering IP\textsubscript{3} production, which moves through diffusion to ER, where opens calcium channels. If the metabotropic receptors deactivate and IP\textsubscript{3} production reduces, the concentration increases more slowly until the outflux from the cell and the return flux to the ER become more relevant, reducing intracellular calcium concentration. These phenomena can be observed on Fig. \ref{fig:4}, which shows concentration profiles after stimulus.

The results obtained can be compared to the ones published in several experimental papers like \citet{venance1997mechanism} and  \citet{ullah2006anti}. However, the experimental conditions used, such as astrocyte cultures and long-term stimulations, in opposition to the situation simulated in this work, from one isolated astrocyte, stimulated only at one of its extremities, can only allow qualitative comparisons.

The diversity of data in the literature, obtained from different cells, regions and experimental methods, makes it impossible to establish precise values for model parameters. In addition, most of them are related to specific local situations. An example is the action of a specific channel that depends not only on its dynamics, but also on its density and, most of times, on its availability in the cell membrane. In this context, it is important to evaluate the relative influence of each parameter of the model on calcium signaling through a parametric sensibility analysis.

\subsection{Parameter sensitivity}
\label{sec:8}

After an analysis of \citet{araque2014gliotransmitters} paper, it is possible to infer that calcium cytosolic average concentration doubles in response to an elementary stimulation. It is still possible to notice that, for this case, signaling time is about 1.0 s. Thus, the choice of model free parameters values aimed that calcium cytosolic average concentration at rest doubled and signalization time was, approximately, 1.0 s.

In the model proposed here there are three free parameters, the constants associated to IP\textsubscript{3} production and degradation, \(K^{p}\) and \(h^{d}\), respectively, and the mass transfer coefficient through IP\textsubscript{3}  activated channels in ER membrane, \(h^{lc}_{ERM}\). Initially is presented a parameter analysis considering the desired values for signaling time and calcium cytosolic average concentration. The deviation, here designed, \(E\), was calculated by the least squares method: \(E=(\overline{y}^{C}_{cyt,max}-2)^2+(t_{C}-1)^2\). In the equation, \(\overline{y}^{C}_{cyt,max}\) is the maximum value reached by dimensionless average concentration of Ca\textsuperscript{2+} cytosolic, and \(t_{C}\) is the time for return of this concentration to the rest value.

\begin{figure*}
  \includegraphics[width=1.0\textwidth]{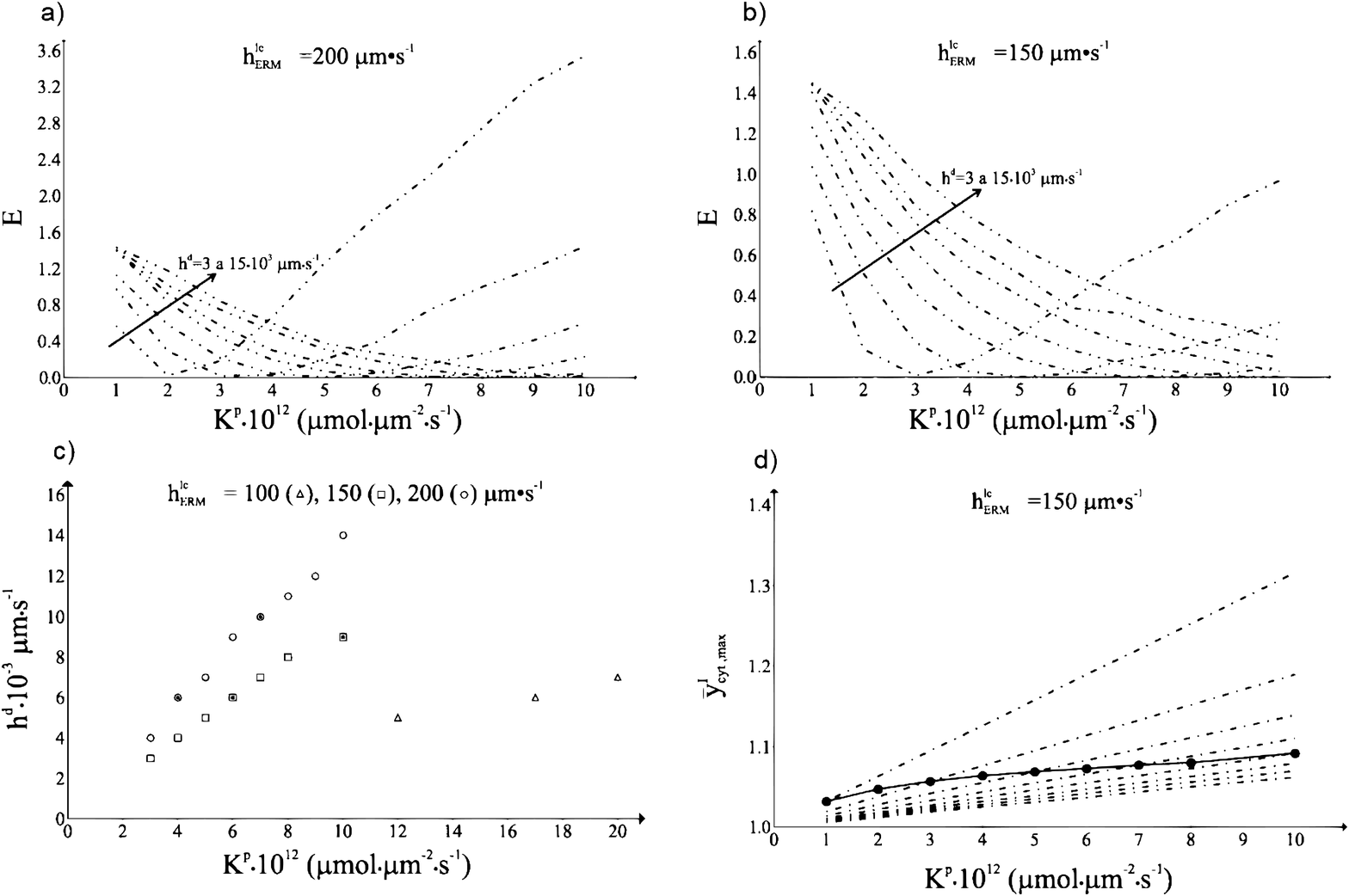}
\caption{\textbf{Effect of three free parameters in values of \(\overline{y}^{C}_{cyt,max}\) and \({t}_{C}\).} \textbf{(a)} Error for  \(h^{lc}_{ERM}=200\) \(\mu m.s^{-1}\). \textbf{(b)} Error for  \(h^{lc}_{ERM}=150\) \(\mu m.s^{-1}\). It may be noted several parameter sets make possible to obtain error values close to zero. Those sets are shown in part \textbf{(c)}. The four highlighted points correspond to lowest error values obtained. The set chosen was \(h^{lc}_{ERM} = 150\) \(\mu m.s^{-1}\), \(h^{d}\) = 6 x \(10^{3} \mu m.s^{-1}\) and \(K^{p}\) = 6 x \(10^{-12}\) \(\mu mol.\mu m.s^{-1}\). \textbf{(d)} Maximum values of IP\textsubscript{3} mean concentration in cytosol for \(h^{lc}_{ERM}=150\) \(\mu m.s^{-1}\). Each dotted line joins points for same value of \(h^{d}\), considering just odd values varying from 1 to 15 x \(10^{3}\). Continuous line joins points with lowest E for \(h^{d}\) ranging from 1 to 15 x \(10^{3}\).
}
\label{fig:5}       
\end{figure*}

Fig. \ref{fig:5} shows the results obtained for \(10^{12}\) x \(K^{p}\in [1;10]\), \(10^{-3}\) x \(h^{d} \in [1;15]\) and \(h^{lc}_{ERM}\) = 100, 150 and 200 \(\mu m.s^{-1}\). Fig. \ref{fig:5}a and Fig. \ref{fig:5}b show \(E\) for two values of \(h^{lc}_{ERM}\), clearly demonstrating its dependency of the free parameters of the model. In Fig. \ref{fig:5}c the indicated sets represent lowers values for  \(E\). For \(h^{lc}_{ERM}=150\) \(\mu m.s^{-1}\), Fig. \ref{fig:5}d shows the maximum values obtained for IP\textsubscript{3} cytosolic average concentration. The continuous line links points to lowers values of \(E\).

It is important to highlight that, for sets containing lower values of \(E\), the IP\textsubscript{3} cytosolic average maximum concentration is restricted to a narrow range. This characteristic is interesting when assessed together with the linear behavior of the relation between the parameters that allow fewer errors.

Considering the results presented at Fig. \ref{fig:5}, the values chosen for the free parameters were: \(K^{p}\) = 6 x \(10^{-12}\) \(\mu mol. \mu m.s^{-1}\), \(h^{d}\) = 6 x \(10^{3}\) \(\mu m.s^{-1}\) and \(h^{lc}_{ERM}=150\) \(\mu m.s^{-1}\).

Now, after the analysis of the parameter set, it will be exposed how each one affects the system behavior.

\subsubsection{IP\textsubscript{3} production}
\label{sec:9}

\begin{figure*}
  \includegraphics[width=1.0\textwidth]{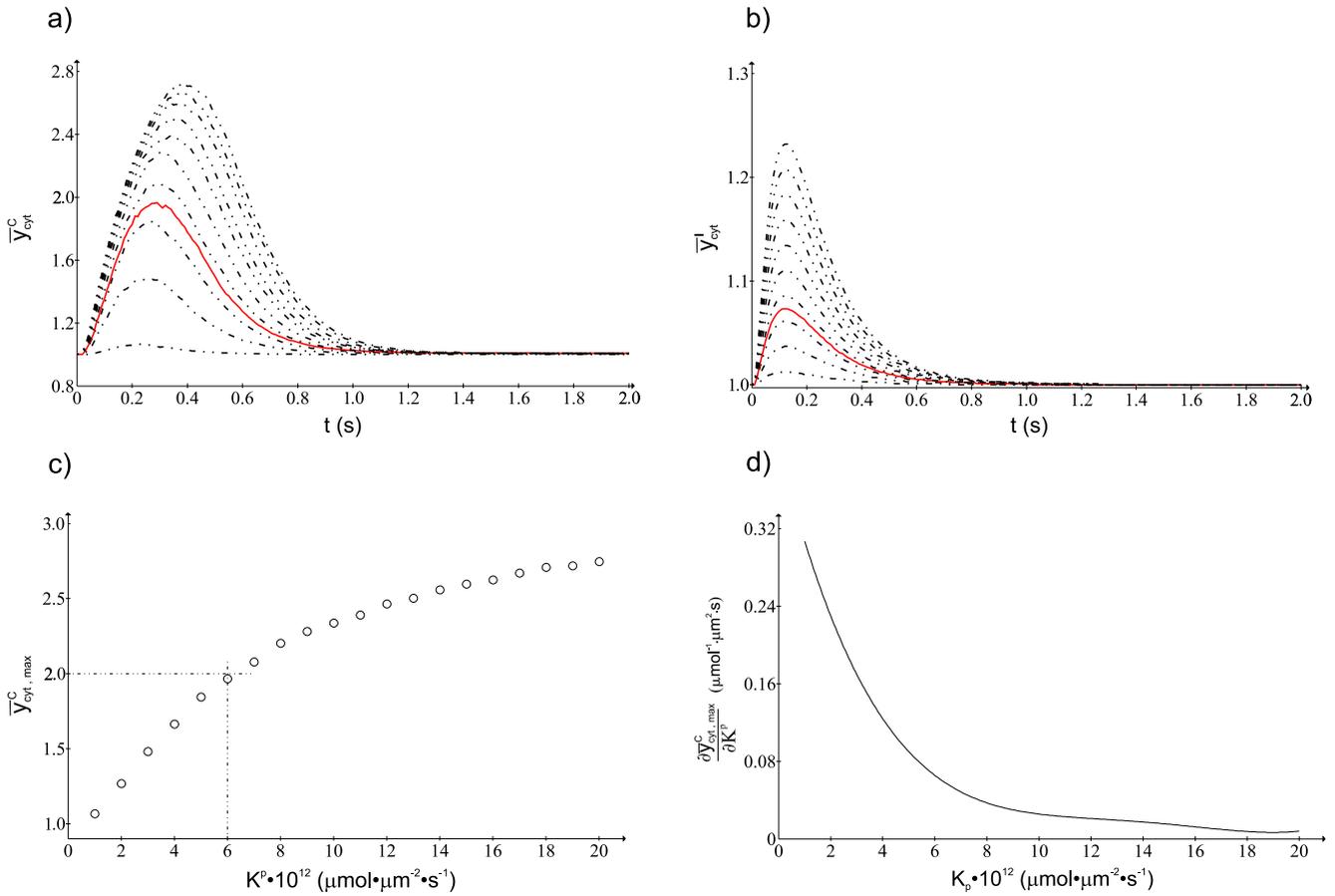}
\caption{\textbf{Dynamics of Ca\textsuperscript{2+} and IP\textsubscript{3} mean concentrations in cytosol as function of global coefficient of IP\textsubscript{3} production rate.} \textbf{a} Dynamics of dimensionless mean concentrations of Ca\textsuperscript{2+} in the cytosol after the glutamatergic stimulation. The dotted lines show the \(\overline{y}_{cyt}^{C}\) dynamics for several values of \(K^{p}\) (1 to 19 x 10\textsuperscript{-12}, considering just the odd values). The continuous line shows the calcium dynamics for \(K^{p}\) = 6 x \(10^{-12} \mu mol. \mu m.s^{-1}\). \textbf{b} Dynamics of dimensionless mean concentrations of IP\textsubscript{3} in the cytosol. The dotted lines show the IP\textsubscript{3} dynamics for odd values of \(K^{p}\), from 1 to 19 x 10\textsuperscript{-12}. The continuous line shows the IP\textsubscript{3} dynamics for \(K^{p}\) = 6 x \(10^{-12}\) \(\mu mol. \mu m.s^{-1}\). For both concentrations, \(\overline{y}_{cyt}^{C}\) and \(\overline{y}_{cyt}^{i}\), the maximum values are directly proportional to \(K^p\). \textbf{c} Maximum values of dimensionless Ca\textsuperscript{2+} concentrations in cytosol for values of \(K^{p}\) from 1 to 20 x 10\textsuperscript{-12}. For \(K^{p}\) = 6 x \(10^{-12}\) \(\mu mol. \mu m.s^{-1}\), the maximum concentration reaches the double of resting concentration. \textbf{d} Derivative of curve of the maximum values of dimensionless Ca\textsuperscript{2+} concentrations  in cytosol in relation to \(K^{p}\).}
\label{fig:6}       
\end{figure*}

The global coefficient of IP\textsubscript{3} production rate affects directly the maximum values of calcium and IP\textsubscript{3} average concentrations in cytosol. Fig. \ref{fig:6}a and Fig. \ref{fig:6}b show temporal profiles of those concentrations. It can be observed that curves pattern and restitution time does not change. Fig. \ref{fig:6}c shows that, as the value of \(K^{p}\) increases, the cytosolic calcium maximum concentration value increases until it reaches a limit, from which any increase in the analyzed parameter does not cause relevant changes in calcium concentration peak.

The  observed behavior in Fig. \ref{fig:6}c is expected, because the increase in IP\textsubscript{3} production results in a higher concentration of this compound in cytosol (Fig. \ref{fig:6}b), which allows the activation of a greater number of channels in ER, releasing more calcium. This release is limited by quantity of channels present in ER membrane, which reduces this compound influence in calcium concentration peak, after specific value. Fig. \ref{fig:6}d shows the derivative graph of maximum calcium curve (Fig. \ref{fig:6}c) in relation to \(K^{p}\). From this last curve, it is possible to conclude the model is more sensitive to parameter \(K^{p}\) in interval of 1 to 10 x 10\textsuperscript{-12} \(\mu mol.\mu m.s^{-1}\). Sensitivity is significantly lower above the last value. The value chosen, \(K^{p}\) = 6 x \(10^{-12}\) \(\mu mol.\mu m.s^{-1}\), is within this region. It is important to show that IP\textsubscript{3} production requires energy expenditure \citep{o2013emerging}, reason to choose values resulting in lower producing for same desired limits of cytosolic calcium.

\subsubsection{IP\textsubscript{3} degradation}
\label{sec:10}
\begin{figure*}
  \includegraphics[width=1.0\textwidth]{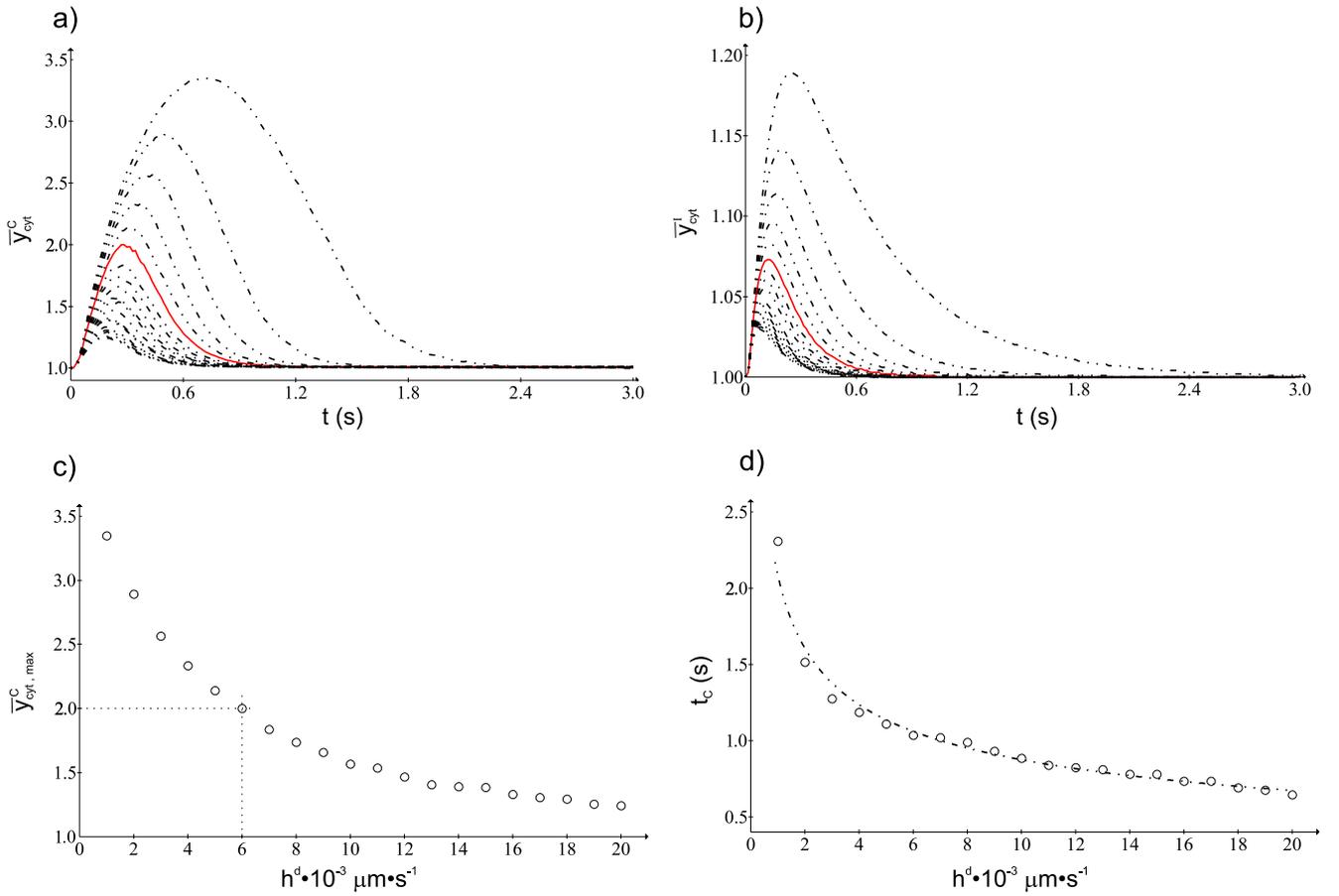}
\caption{\textbf{Dynamics of Ca\textsuperscript{2+} and IP\textsubscript{3} mean concentrations as function of IP\textsubscript{3} degradation coefficient.} Dynamics of \textbf{a} Ca\textsuperscript{2+} and \textbf{b} IP\textsubscript{3} dimensionless mean concentrations in cytosol for odd values of \(10^{-3}\) x \(h^{d}\) between 0 and 20. For both concentrations, \(\overline{y}_{cyt}^{C}\) and \(\overline{y}_{cyt}^{i}\), the maximum values are inversely proportional to \(h^d\). \textbf{c} Maximum values and \textbf{d} restitution time for cytosolic Ca\textsuperscript{2+} mean concentration for values of \(10^{-3}\) x \(h^{d}\) from 1 to 20.}
\label{fig:7}       
\end{figure*}

The IP\textsubscript{3} degradation, as its production, affects cytosolic calcium concentration peak, as shown in Fig. \ref{fig:7}a and Fig. \ref{fig:7}c. How much higher is the parameter associated to degradation, lower is the value reached by calcium concentration. Fig. \ref{fig:7}d shows restitution time for cytosolic Ca\textsuperscript{2+} average concentration. Observed restitution time varies significantly for low values of \(h^{d}\), reaching a region with small variations for higher values. The beginning of this region can be defined in \(h^{d}=6000\) \(\mu m.s^{-1}\). As shown in Fig. \ref{fig:7}c, at this point, maximum average concentration reaches twice the rest value. That is the chosen value, because it reduces restitution time and maximizes cytosolic Ca\textsuperscript{2+} maximum concentration.

\subsubsection{Ca\textsuperscript{2+} release}
\label{sec:11}

\begin{figure*}
  \includegraphics[width=1.0\textwidth]{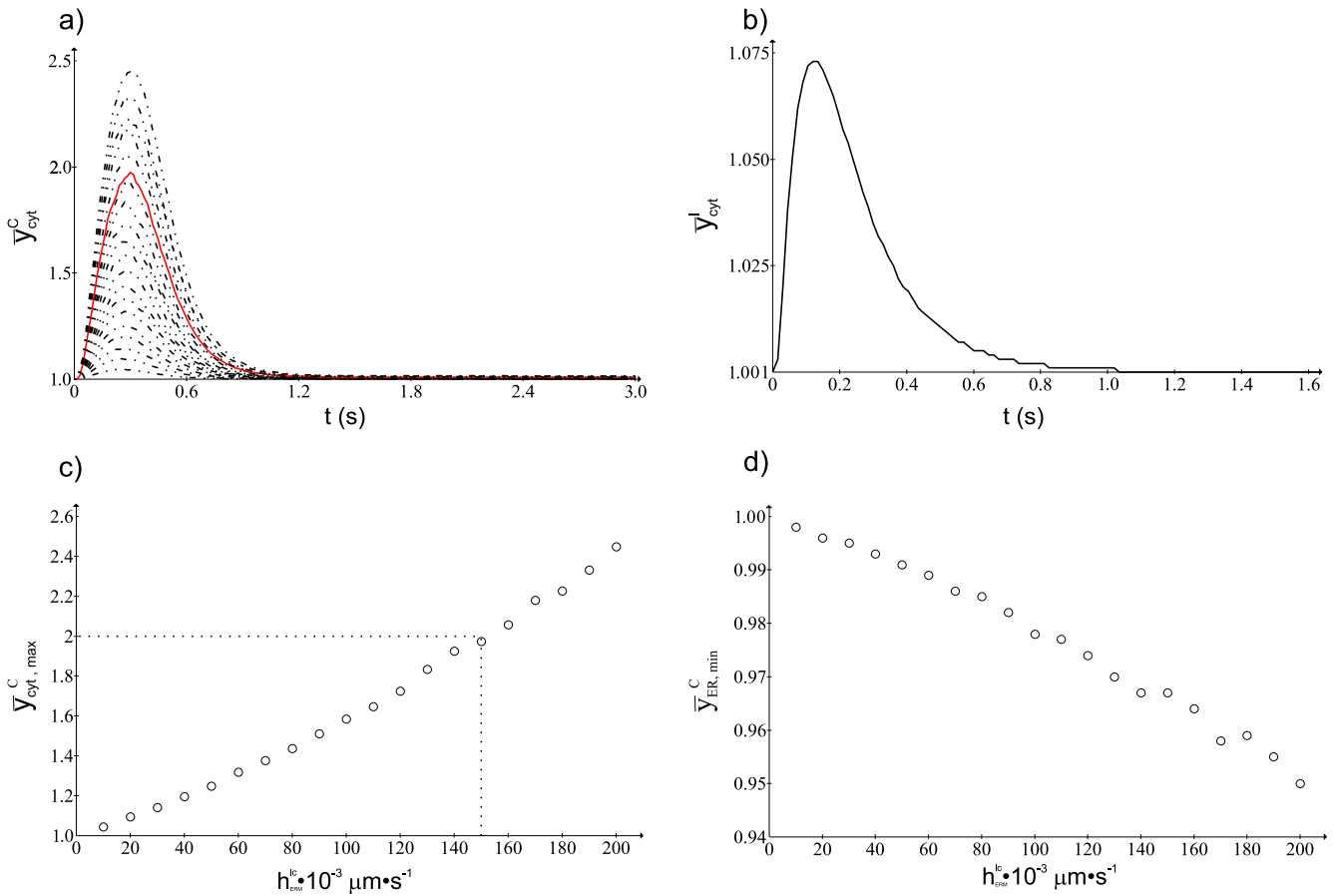}
\caption{\textbf{Dynamics of dimensionless Ca\textsuperscript{2+} mean concentrations in cytosol and in ER as a function of mass transfer coefficient of ligand channels of ER membrane.} \textbf{(a)} Dynamics of Ca\textsuperscript{2+} mean concentration in cytosol for \(10^{-3}\) x \(h^{lc}_{ERM} \in [10;200]\). The continuous line correspond to \(10^{-3}\) x \(h^{lc}_{ERM} = 150 \mu m.s^{-1}\). \textbf{(b)} Dynamics of IP\textsubscript{3} mean concentration in cytosol. This dynamics does not depends on the values of  \(h^{lc}_{ERM}\). \textbf{(c)} Maximum values in cytosol and \textbf{(d)} minimum values in ER for dimensionless Ca\textsuperscript{2+} concentration.}
\label{fig:8}       
\end{figure*}

Calcium release from endoplasmic reticulum depends on mass transfer coefficient through ligand dependent channels, \(h^{lc}_{ERM}\). Fig. \ref{fig:8}a shows its influence in cytosolic Ca\textsuperscript{2+} average concentration evolution. Fig. \ref{fig:8}c shows maximum values of this variable for different values of \(h^{lc}_{ERM}\), evidencing calcium release is directly proportional to this parameter. The constant growth observed at Fig. \ref{fig:8}c is justified by a higher mass transfer coefficient, producing more calcium exit from ER to cytosol. As the process is limited by ER concentration, which is significantly high, no  saturation were observed in analyzed interval. Also IP\textsubscript{3} cytosolic average concentration is not affected by this parameter (Fig. \ref{fig:8}b). Fig. \ref{fig:8}d shows little influence of it in relation to ER Ca\textsuperscript{2+} minimum concentration. It is important to mention that calcium concentration in ER reduces about 4.0\% when in cytosol increases 100\%. The value chosen for the parameter, \(150\) \(\mu m.s^{-1}\), is, again, associated to desired calcium maximum cytosolic concentration (\(0.2\) \(\mu M\)).

\subsection{Implication of TRPC incorporation}
\label{sec:12}

\begin{figure*}
  \includegraphics[width=1.0\textwidth]{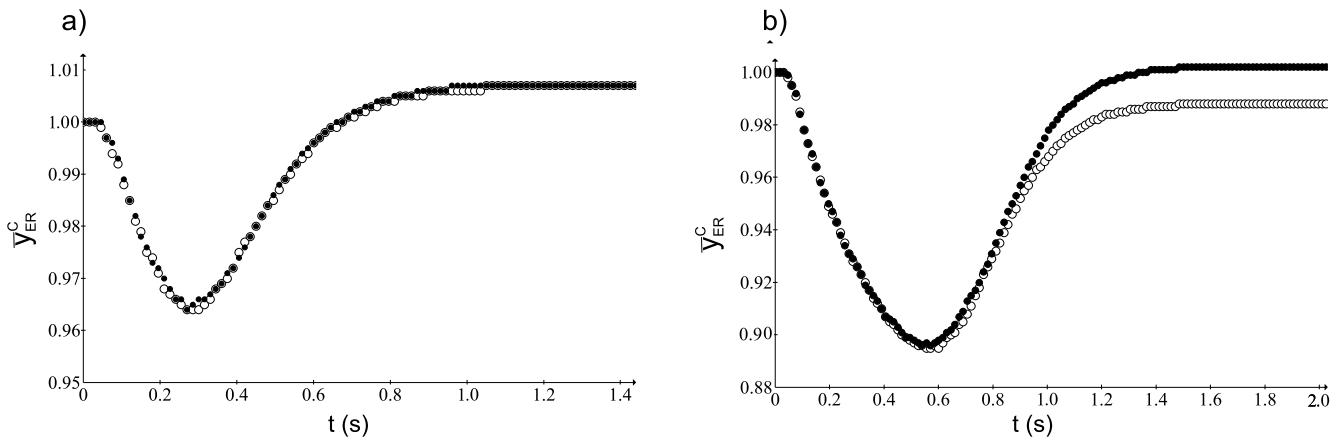}
\caption{\textbf{Comparison between dynamics of calcium concentration in ER, obtained with and without the activation of TRPC channels.} In the figure, the white circles represent the dynamic without the activation of the TRPC channels. The black circles represent the activation of this mechanism. \textbf{a} Curves obtained in the case of one single stimulation, with 0.01 s duration. \textbf{b} Curves obtained in the case of a burst, with a series of 10 stimulations of 0.01 s each one.}
\label{fig:9}       
\end{figure*}

Incorporation of the TRPC, while not altering cellular dynamics, helps Ca\textsuperscript{2+} concentration in ER to return to resting value in a reasonable time interval \citep{malli20072+,verkhratsky2014store}. However, as shown in Fig. \ref{fig:9}, the effect is not significant in case of a single glutamatergic stimulation lasting 0.01 s. By contrast, in case of burst, with 10 stimulations in sequence, TRPC incorporation allows calcium baseline concentration returning to rest value in ER. It means this incorporation is important for an astrocyte immersed in cell medium, with several connections and submitted to several stimuli.

\subsection{Opening probability}
\label{sec:13}

\begin{figure*}
  \includegraphics[width=1.0\textwidth]{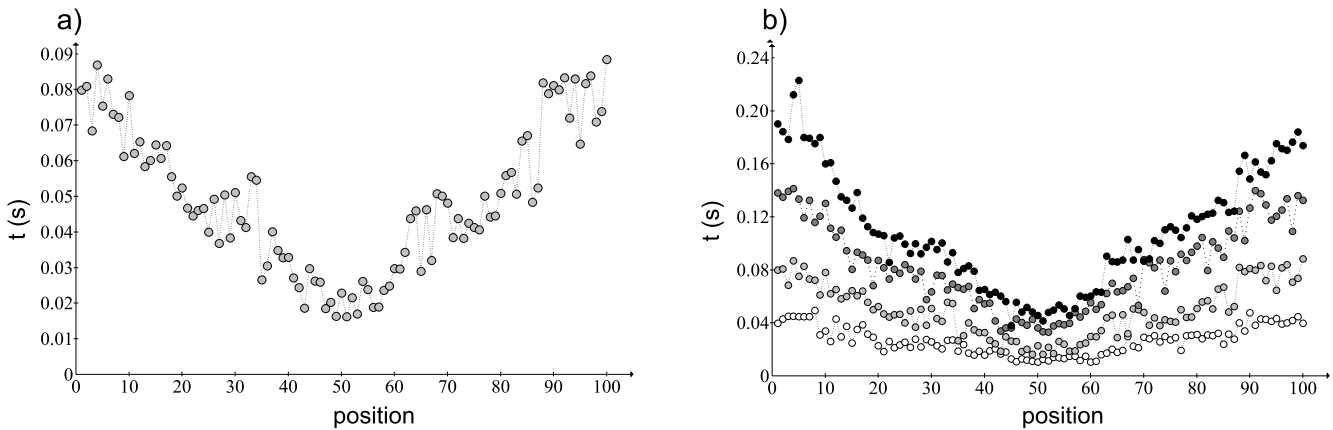}
\caption{\textbf{IP\textsubscript{3} activated channel opening time in different points in the contour of ER membrane.} \textbf{a} Opening time for all ER channels, with parameter values indicated at Sect.~\ref{sec:8}. \textbf{b} Opening time for the same channels, for several values of \(n_{I}\). The values of this parameter, from the lighter to darker points are 1.0, 2.2, 5.0 and 10.0, respectively. As the parameter value increases, the opening probability reduces and the channels take more time to open.}
\label{fig:10}       
\end{figure*}

The use of \(P_{I}\) probability (Eq. \ref{Equation 10}) for opening IP\textsubscript{3}-activated channels in endoplasmic reticulum membranes requires their individualization and numbering. There are one hundred equally spaced channel positions distributed along the ER membrane. Fig. \ref{fig:3} shows numbering starting in opposite face and going around membrane clockwise. When occur stimulus in a synapse connected to an astrocyte arm, the ER side facing that arm has channels with positions 38 to 62, 50 being the central position. On the opposite side, the channel in central position and farthest from stimulation has number 100. The channels do not open in sequence but obey the probability test. It is possible to see at Fig. \ref{fig:10}a opening times for all channels, when parameter values are those described on Section \ref{sec:8}. An analysis of parameters influence at Eq. \ref{Equation 8}, \(n_{I}\), \(n_{C}\), \(ks_{C}\), and \(Ks_{C}\), shows that \(P_{I}\) is mainly affected by the first one. At Fig. \ref{fig:10}b it is possible see this effect. The values used for exponent \(n_{I}\) were 1.0, 2.2, 5.0 and 10.0. The parameter reduces probability and increases opening time.

\subsection{Influence of ER in branches}
\label{sec:14}

ER is found in astrocyte branches \citep{golovina2000unloading, golovina2005visualization,holtzclaw2002astrocytes}. However, its contribution to calcium dynamics is still unclear \citep{oheim2017local}. In order to study this influence, the computational program used in this work was modified to allow the allocation of reticulum in the branches.

Initially, calcium dynamics was simulated in only one branch to verify how reticulum would influence it. The branch had 6 \(\mu m \) length and 1 \(\mu m \) width. The ER therein was placed at a distance of 0.4 \(\mu m \) from the astrocyte membrane around the synapse and had 1 \(\mu m \) length. Other lengths were tested as well: 1.5 \(\mu m \), 2.0 \(\mu m \), 2.5 \(\mu m \), and 3.0 \(\mu m \).

\begin{figure}
  \includegraphics[width=0.48\textwidth]{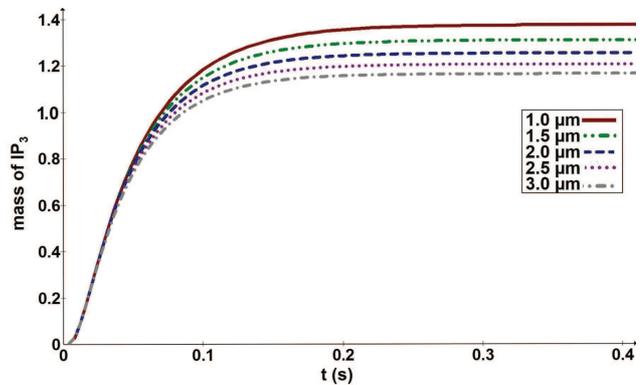}
\caption{\textbf{Effect of ER length in the mass of IP\textsubscript{3} leaving the astrocyte branch.}  The curves show the mass sum over time that leaves the astrocyte branching for each reticulum size evaluated. }
\label{fig:11}       
\end{figure}

First we analyzed the ER effect in astrocyte branch in IP\textsubscript{3} dynamics, because this is mainly responsible for propagation of calcium signaling. It was observed that ER decreases the IP\textsubscript{3} mass that leaves the branch, as can be seen in Fig. \ref{fig:11}. All ER sizes cause this same effect; however, the larger the endoplasmic reticulum size, the smaller the amount of IP\textsubscript{3} leaving the branching. This is due to the IP\textsubscript{3} degradation phenomenon that occurs in ER contours. The larger the endoplasmic reticulum, the greater the degradation. Consequently, the mass of IP\textsubscript{3} available to go ahead in the cell and to release calcium from stocks is lower.

As showed in Fig. \ref{fig:11}, endoplasmic reticulum allocation to astrocyte branch may not be good for global Ca\textsuperscript{2+} signaling, because the ER decreases the IP\textsubscript{3} available amount for opening calcium channels in cell body ER. Thus, it is necessary to determine why cells would have ER in astrocyte branchs.

\begin{figure}
  \includegraphics[width=0.48\textwidth]{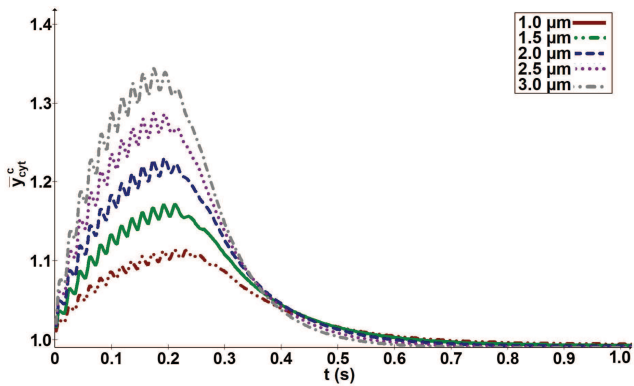}
\caption{\textbf{Effect of ER length in the dimensionless Ca\textsuperscript{+2} concentration in astrocyte branch.}  The curves show the dimensionless Ca\textsuperscript{+2} concentration over time to each reticulum size evaluated. }
\label{fig:12}       
\end{figure}

Calcium has local signals not spread to whole cell \citep{ di2011local,kanemaru2014vivo}. These signals are short and fast. The endoplasmic reticulum may contribute to them. Fig. \ref{fig:12} shows the dynamics of dimensionless calcium concentration in branches. From this figure, it is possible to conclude that ER promotes local calcium signaling, increasing the levels of Ca\textsuperscript{2+} in a smaller time interval.

Calcium dynamics observed in Fig. \ref{fig:12} can be compared to experimental results of \citet{ di2011local} with respect to local calcium events. Studying astrocyte branches, researchers concluded that local signaling extends at distances of 3 to 8 \(\mu m\), having 0.14 s rise time and 0.7 s duration. In this work, the branch has 6 \(\mu m\), calcium concentration rise time was 0.2 s and duration was 0.8 s. These results can validate the parameters set defined in the previous simulation.

\begin{figure}
  \includegraphics[width=0.48\textwidth]{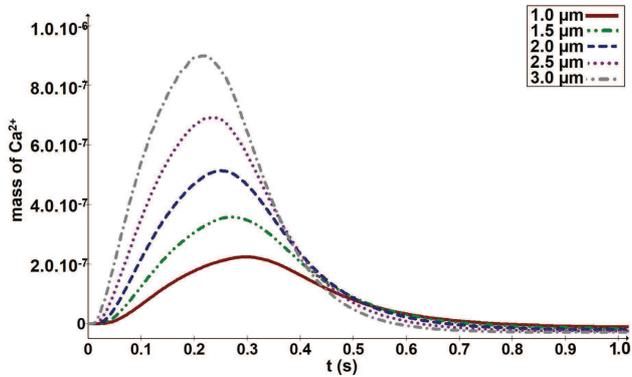}
\caption{\textbf{Effect of ER length in the mass of Ca\textsuperscript{+2} leaving the astrocyte branch.}  The curves show the mass of Ca\textsuperscript{+2} that leaves the astrocyte branch over time to each reticulum size evaluated. }
\label{fig:13}       
\end{figure}

In addition to contributing to local calcium signaling, ER could contribute to calcium wave advancement, because the released calcium go ahead in the cell when reaches the astrocyte branch end. As can be seen in Fig. \ref{fig:11} and Fig. \ref{fig:13}, the amount of calcium leaving the branching is smaller and slower than IP\textsubscript{3}; therefore, this calcium possibly does not contribute significantly to calcium wave advancement.

The production of IP\textsubscript{3} depends on calcium concentration surrounding metabotropic receptors. Another effect of ER on astrocyte branch endfeet could be enhancing IP\textsubscript{3} production. In the cases discussed above, this effect is imperceptible, because the time for IP\textsubscript{3} production is shorter than time IP\textsubscript{3} takes to reach endoplasmic reticulum to release calcium and for calcium to arrive in the membrane to strengthen IP\textsubscript{3} production. It was necessary to evaluate a burst corresponding to successive synaptic transmissions. Three stimuli were performed in order to determine whether the calcium released in the first stimulus was able to influence IP\textsubscript{3} production in second and third stimuli. An insignificant effect was found. IP\textsubscript{3} production increases by about 0.001\%.

In addition to reticulum size, endoplasmic reticulum position may influence IP\textsubscript{3} production. To evaluate the ER position influence on IP\textsubscript{3} production, the smallest reticulum tested (1.0 \(\mu m\)) was allocated at 0.4 \(\mu m\), 0.8 \(\mu m\) and 1.4 \(\mu m\) away from cell membrane where IP\textsubscript{3} is produced. Again, the effect on IP\textsubscript{3} production was less than 0.001\%.

With this considerations it is possible to verify that ER allocated in astrocyte branches mainly contributes to local calcium signaling. In relation to global signaling, its effect is negative because it reduces amounts of IP\textsubscript{3} available for channel opening in cell body ER. It was also possible to note that increasing calcium local concentration does not significantly influence IP\textsubscript{3} production.

\begin{figure}
  \includegraphics[width=0.48\textwidth]{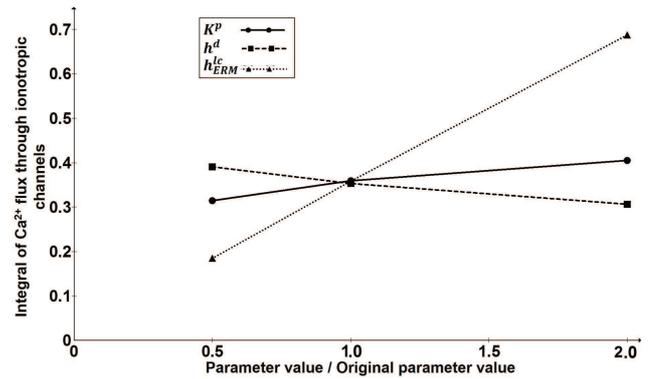}
\caption{\textbf{Effect of the parameters in the calcium flux through the channels coupled to ionotropic receptors.}  Influence of the parameters related to the production of IP\textsubscript{3}, degradation of IP\textsubscript{3} and release of calcium in the calcium flux through the channels coupled to ionotropic IP\textsubscript{3} receptors present in the endoplasmic reticulum. The original parameters values are \(K^{p}\) = 6.0 x \(10^{-12} \mu mol\), \(\mu m^{-2}\)\(s^{-1}\), \(h^{d}\) = 6.0 x \(10^{3}\) \(\mu m\)\(s^{-1}\), \(h^{lc}_{ERM}\) = 150 \(\mu m\)\(s^{-1}\).}
\label{fig:14}       
\end{figure}

To analyze parameters influence in this new simulation, the same parameters from previous simulation were evaluated: production (\(K^{p}\)) coefficient, degradation (\(h^{d}\)) coefficient and calcium release (\(h^{lc}_{ERM}\)) coefficient. The results can be seen in Fig. \ref{fig:14}.

As previously discussed, parameters related to IP\textsubscript{3} have opposite effects: the higher the production coefficient, the greater the calcium amount released by ER. The higher the degradation coefficient, the less calcium is released from ER. Fig. \ref{fig:14} shows that the influence of these parametrs is not significant when evaluating the flux through channels coupled to ionotropic IP\textsubscript{3} receptors. Regarding the calcium release parameter, the model sensitivity is much higher. This is because it is directly related to calcium output flux. If the parameter doubles, calcium output flux doubles. If the parameter is halved, the output flux is reduced by half as well.

\begin{figure}
  \includegraphics[width=0.48\textwidth]{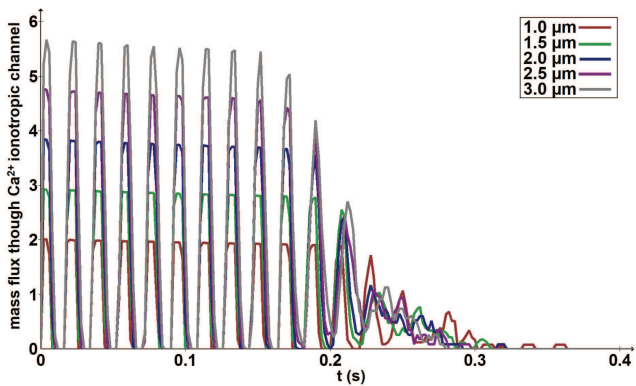}
\caption{\textbf{Calcium flux through the calcium channels coupled to ionotropic receptors of IP\textsubscript{3}.}  The channels coupled to the ionotropic receptors of IP\textsubscript{3} open simultaneously in the endoplasmic reticulum. The larger the reticulum, the greater the number of channels and, consequently, the greater the outflux of calcium. With the passage of time, the number of channels that opens is smaller and the effect of the probability of opening becomes evident.}
\label{fig:15}       
\end{figure}

Another interesting aspect to emphasize is channel opening process. In the cell body ER, IP\textsubscript{3} arrives slowly, so channels are opening sequentially (Fig. \ref{fig:10}). In branching, by contrast, given the high IP\textsubscript{3} diffusion coefficient and the small space for movement, channels open almost simultaneously. This phenomenon is evidenced in Fig. \ref{fig:15}. For all ER sizes, all channels are opened simultaneously when the concentration of  IP\textsubscript{3} is high. When IP\textsubscript{3} concentration decreases around 0.2 s, it is possible to observe more clearly the opening probability effect in the channels.

\begin{figure}
  \includegraphics[width=0.48\textwidth]{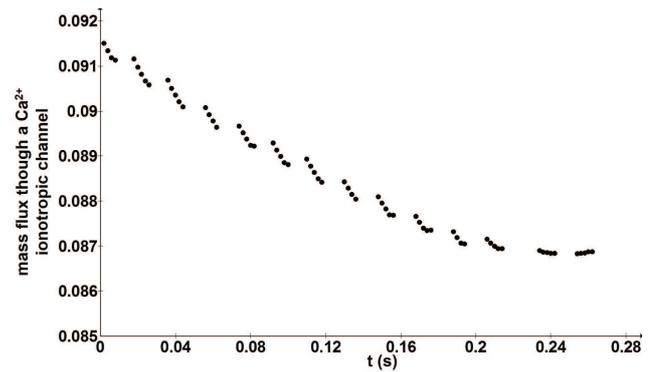}
\caption{\textbf{Calcium flux through a single channel coupled to an ionotropic IP\textsubscript{3} receptor.}  The flux of calcium through a channel, for multiple stimuli, decreases over time due to decreased gradients. The blank intervals represent the periods of time that the channel remains closed.}
\label{fig:16}       
\end{figure}

It is also possible to observe in Fig. \ref{fig:15} small decreases in output flux over time despite the fact that all channels open simultaneously. This can be best understood by observing flux over time through a single channel (Fig. \ref{fig:16}). As the gradients diminish (increasing calcium concentration in the cell and decreasing calcium concentration in the ER), the flux also decreases.

\begin{figure}
  \centering
  \includegraphics[width=0.25\textwidth]{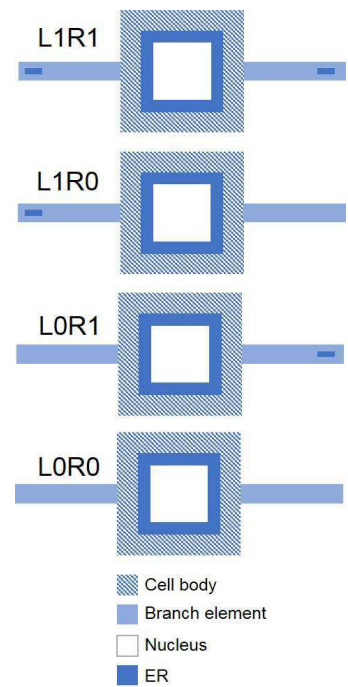}
\caption{\textbf{Geometric configurations implemented computationally in way to analyze endoplasmic reticulum effects on astrocytes branches.} The figure shows the four configurations used. From top to bottom, L1R1 has two endoplasmic reticulum, in left and right branches; L1R0 only has the reticulum in the left branch; L0R1 only in the right branch; L0R0 has no endoplasmic reticulum in the branches. The results generated by these four configurations were analyzed and compared. The length of the endoplasmic reticulum is 1 \(\mu m\).}
\label{fig:17}       
\end{figure}

Computational simulation of an astrocyte branch allowed analysis of ER effects on local calcium signaling. Analysis of its effects on global calcium signaling requires a more complex geometry. Therefore, we simulateed the geometry shown in Fig. \ref{fig:17}, containing a cell body and two branches connected to it. The two branches contain ER. This configuration was named L1R1 (one ER at left branch and one ER at right branch) to facilitate results presentation. For better result analysis, we tested configurations with the ER in only one of the branches (L1R0 and L0R1). Finally, for comparison purposes, we also evaluate the situation with branches without reticulum (L0R0).

\begin{figure}
  \includegraphics[width=0.48\textwidth]{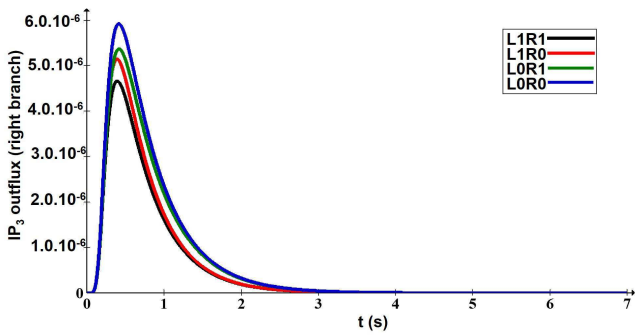}
\caption{\textbf{Instantaneous IP\textsubscript{3} outflux at the end of the right branch.} The graph shows the outflux of IP\textsubscript{3} on the branch opposite the branch where the stimulus is applied. The flux is smaller for the situation where there are ER in the two branches, intermediate for the situations with ER in only one of the branches and greater for the case in which the branches do not have ER.}
\label{fig:18}       
\end{figure}

First we evaluated branches endoplasmic reticulum effect on IP\textsubscript{3} production. It has been observed that configurations L1R1 and L1R0 have 0.000086\% greater IP\textsubscript{3} production than L0R1 and L0R0. This shows that reticulum located near stimulus increases IP\textsubscript{3} production. Although the increase is not significant, it does occur. This IP\textsubscript{3} production increase occurs because calcium released from reticulum by IP\textsubscript{3} travels to cell membrane and therein can potentiate IP\textsubscript{3} production \citep{pawelczyk1997structural}. This can be evidenced when the first stimulus does not cause distinction among fluxes produced in the four cases. This distinction can be noticed in subsequent stimuli (results not shown). What makes so small IP\textsubscript{3} production increase is the slow temporal calcium dynamics. Thus, it is possible to state that reticulum at branch ends does not significantly contribute to increase IP\textsubscript{3} production.

With respect to IP\textsubscript{3} output flux, that is, to IP\textsubscript{3} that would advance in the cell or even in astrocytes network, it is possible to conclude that the most favorable configuration would be the one in which there is no endoplasmic reticulum in the branches (Fig. \ref{fig:18}). This is because the ER promotes this compound degradation. The IP\textsubscript{3} degradation will be proportional to endoplasmic reticulum amount in cell.

The L1R0 and L0R1  configurations have different IP\textsubscript{3} flux profiles at the right branch end. The reason is that, at the stimulated end, L1R0 configuration, a large concentration of IP\textsubscript{3} reaches the ER and is degraded. The degradation of IP\textsubscript{3} is proportional to its concentration in the cytosol, close to the ER membrane. In the L0R1 configuration, the IP\textsubscript{3} flow that reaches the last ER is diluted by diffusion and because it has already been partially degraded in the central ER. The lower concentration in the proximity of the last ER membrane leads to less degradation of IP\textsubscript{3} and a greater amount reaches the outlet at the end of the right branch.

\begin{figure}
  \includegraphics[width=0.48\textwidth]{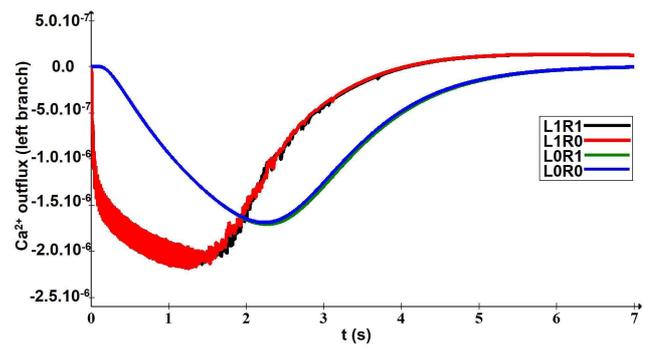}
\caption{\textbf{Instantaneous calcium flux in the cell membrane where stimulus occurs.} Calcium flux in left end membrane of the branch where the stimulus occurs is outflux, so it is negative. When there is ER in this branch, the flux is larger and more oscillatory. This oscillation is due to probabilistic nature of channels opening. When there is no ER, diffusion dissipates this probabilistic effect, no longer being perceived.}
\label{fig:19}       
\end{figure}

Regarding calcium, it is interesting to note calcium flux in cell membrane in which the stimulus occurs, showed in Fig. \ref{fig:19}. In this region, there is a long time outflux, because increase of intracellular calcium concentration promotes pump action attempting to restore the resting state. In cases when ER is in the left branch, outflux occurs faster and with great oscillation. This oscillation is result of probabilistic dynamics of channel opening in the reticulum. As the ER is close to where flux is being measured, this probabilistic effect is more clearly perceived. In cases where there is no ER in vicinity of the membrane where stimulus occurs, calcium that arrives there comes from cell body. As it needs to travel through cell body and branch to leave the cell, the probabilistic effect is no longer evident.

\begin{figure}
  \includegraphics[width=0.48\textwidth]{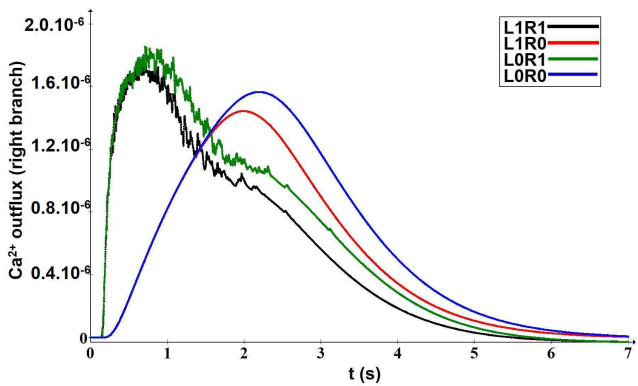}
\caption{\textbf{Instantaneous calcium flux at right branch outlet.} The fastest dynamics correspond to settings where there is right branch endoplasmic reticulum, L1R1 and L0R1. The differences between configurations with or without left branch reticulum are due to IP\textsubscript{3} degradation by this ER, reducing IP\textsubscript{3} amount in ER of cell body. The conclusion is that ER in branches away from stimulus sites accelerates calcium advancement in one individual astrocyte, or in a network of them.} 
\label{fig:20}       
\end{figure}

The calcium outflux in the branch end where the stimulus does not occur, right branch, can be observed in Fig. \ref{fig:20}. This figure shows two types of behavior. The first behavior corresponds to configurations containing endoplasmic reticulum in the right branch. The second behavior corresponds to configurations without reticulum in this branch. The ER close to output causes output stream to start earlier. This is because IP\textsubscript{3} quickly hits this ER, releasing calcium. When there is no reticulum, calcium leaving the branch end comes from the ER of cell body. As calcium diffusion is a tenth of IP\textsubscript{3} diffusion, it takes longer to come out in this configuration.

\begin{figure}
  \includegraphics[width=0.48\textwidth]{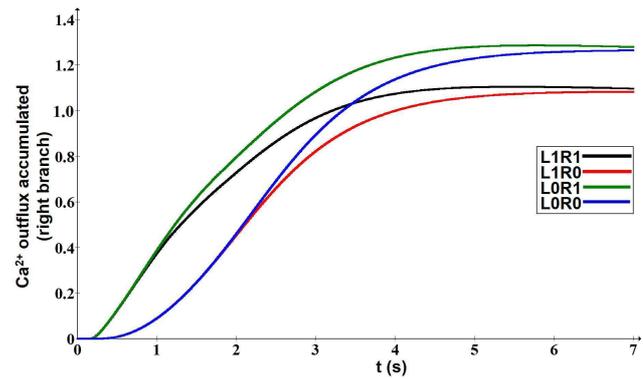}
\caption{\textbf{Accumulated calcium flux at right branch outlet.} The accumulated calcium flux at the astrocyte outlet can be seen in two aspects, quantity and velocity. In all configurations IP\textsubscript{3} amount produced by glutamatergic stimulus is the same. Existence of endoplasmic reticulum in the branch near the stimulus, L1R1 and L1R0 configurations, causes IP\textsubscript{3} degradation, reducing the amount that reaches the ER in central boddy of the cell. This decreases calcium total amount that leaves the astrocyte at the stimulus opposite end. In cell and reticulum here used dimensions, this reduction is approximately 14.5\%. By contrast, when observing velocity, reticulum existence at the exit end, L1R1 and L0R1 configurations, increases the leaving calcium flux velocity. The reason is clear: IP\textsubscript{3}'s higher rate of diffusion allows it to reach the last ER and release calcium before the arrival of calcium generated in central ER.}
\label{fig:21}       
\end{figure}

When calculating the accumulated flux over time, the sequence of calcium outflux value is L0R1$>$L0R0$>$ L1R1$>$L1R0. Although calcium outflux occurs quicker when ER is near of the exit, the total calcium amount leaving the cell is larger when there is no ER in the other branch (Fig. \ref{fig:21}). This is because the left branch reticulum, in addition to releasing calcium, promotes IP\textsubscript{3} degradation. In this way, IP\textsubscript{3} available to reach right branch reticulum is smaller, releasing less calcium. From these results, it is concluded that the configuration allowing greater amount of calcium advance is the one without ER in the first branch. This ER incorporation reduces about 17\% of calcium amount that would advance either in astrocyte network or in cell itself. Now, with respect to the ER near to the exit, it increases only 1\% the amount of calcium coming out.

There are two ways to analyze the effect of endoplasmic reticulum positioned in the branches of astrocytes: one with respect to the amount of calcium and IP\textsubscript{3} available to advance in the cell itself or in a network of astrocytes, and the other with respect to the speed of progress of these compounds. The positioning of ER in the branches decreases the amount of IP\textsubscript{3} that advances in the cell. As a result, there is less calcium release in the ERs that are further ahead. By contrast, the ER positioned in the branches ahead makes the calcium dynamics faster, because IP\textsubscript{3}, moving quickly, reaches them earlier, releasing calcium at the site.

\begin{figure}
  \includegraphics[width=0.48\textwidth]{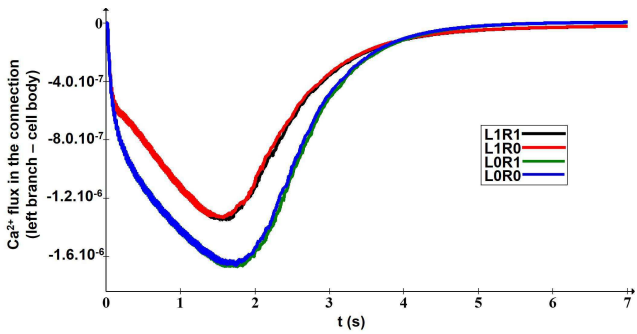}
\caption{\textbf{Flux through connection between left branch and cell body.} The lines are coincident, differentiated only by the existence or not of endoplasmic reticulum in left branch. L1R1 and L1R0 configurations have a lower calcium retrograde motion than L0R1 and L0R0 configurations. This is possibly for two reasons, whose effects add up. The ER in the left branch degrades IP\textsubscript{3} causing smaller amounts to reach central ER and can at the same time produce a calcium barrier that reduces calcium return by diffusion.}
\label{fig:22}       
\end{figure}

One hypothesis about organelles location and function such as  ER in astrocytes branches involve diffusion barriers formation \citep{oheim2017local}. It is possible to evaluate the calcium barrier effect when analyzing flux in connection between branches and cell body. Fig. \ref{fig:22} shows flux through connection between left branch and cell body. To understand the graph is important to consider that positive fluxes indicate left to right and negative fluxes indicate right to left direction. Because this flux is negative, the calcium that passes through the connection comes from the ER of cell body. This happens because IP\textsubscript{3} quickly gets there and the central ER is closer to the connection than the branch ER. Comparing the L1R0 and L0R0 configurations, red and blue lines, respectively, it is possible to visualize the calcium barrier, because the calcium flux towards the branch decreases considerably.

\begin{figure}
  \includegraphics[width=0.48\textwidth]{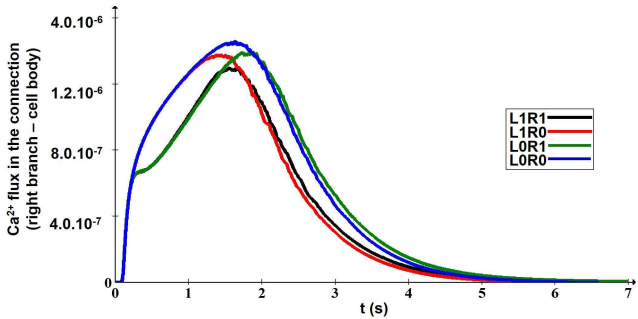}
\caption{\textbf{Flux through connection between right branch and cell body.} Black and green lines show calcium barrier effect. IP\textsubscript{3} reaches the right branch ER and releases calcium. Calcium moves by diffusion in both directions. Movement in retrograde direction (right branch to cell) obstructs the front advancement of calcium originating in central ER. This effect is observed at approximately 0.2s and does not exist in configurations represented by blue and red lines. The effect of left branch ER is highlighted in black and red lines. Calcium flux reaches a smaller total amount and ends earlier. The reason is that left branch ER degrades IP\textsubscript{3}, reducing the amount that reaches central ER. Finally, blue and green lines differ only by calcium barrier effect produced by right branch ER.
}
\label{fig:23}       
\end{figure}

Given the calcium barrier existence, the flux through the connection between cell body and right branch would be expected to be higher in L1R0 situation than flux in L0R0 configuration. However, the flux in L0R0 is higher (Fig. \ref{fig:23}). This occurs because less IP\textsubscript{3} is made available to cell body when there is reticulum in left branch. 

\begin{figure}
  \includegraphics[width=0.48\textwidth]{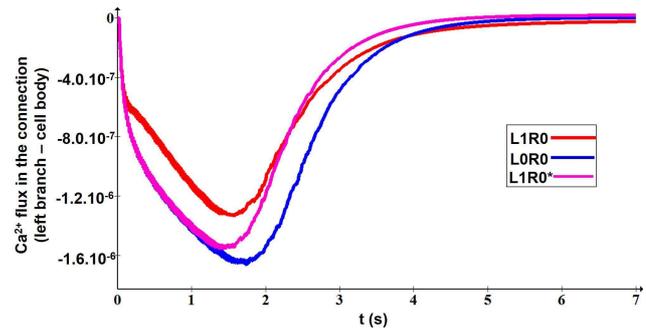}
\caption{\textbf{Flux through connection between left branch and cell body compared to the situation where the left branch ER does not release calcium.} The red and pink lines show the flow of calcium in situations in which the left branch ER respectively releases or not calcium. It is possible to observe the calcium barrier effect. In situation L1R0 the flow is smaller. Red and blue lines difference reveals the effect in calcium flow by left branch ER IP\textsubscript{3} degradation. In left branch ER absence more IP\textsubscript{3} reaches the cell body ER causing greater calcium release.}
\label{fig:24}       
\end{figure}

To verify if calcium barrier really causes the flow decrease in the connection between left branch and cell body, the L1R0 simulation was performed, but the ER only degraded IP\textsubscript{3} and did not release calcium. This simulation was called L1R0*. As can be seen in Fig. \ref{fig:24}, the difference between the blue (L0R0) and pink (L1R0*) curves shows that IP\textsubscript{3} degradation actually decreases this flow. By contrast, the difference between pink (L1R0*) and red (L1R0) shows the calcium barrier effect on this flow.

\begin{figure}
 \includegraphics[width=0.48\textwidth]{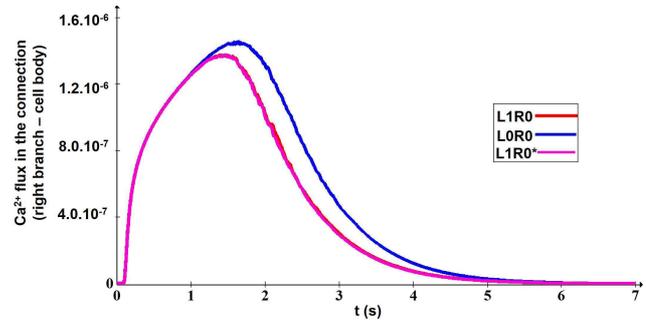}
\caption{\textbf{Flux through connection between right branch and cell body compared to the situation where the left branch reticulum does not release calcium.} The graph shows that while there may be a calcium barrier formed by the left branch ER calcium release, its effects are dissipated along the calcium pathway. Therefore, in the connection between right branch and cell body, there is no distinction between pink and red lines. At this point, the only effect observed on calcium flow is IP\textsubscript{3} degradation promoted by the left branch ER.}
\label{fig:25}       
\end{figure}

In order to evaluate if calcium barrier promotes calcium wave advancement, one can observe Fig. \ref{fig:25}, which compares calcium fluxes in the connection between right branch and cellular body. It is possible to observe no distinction between L1R0 and L1R0* configurations. This shows the calcium barrier effects are dissipated along the calcium pathway. In this way, the barrier contributes to local, but not global, dynamics.

Based on these results, it is possible to notice that several processes influence calcium signaling but there is a clear distinction between local and global effects \citep{di2011local}, as well as between dynamic and quantitative effects. ER on branches participates primarily in the local effects and contributes mainly to the local process dynamics. The ER, when disposed in vicinity of where the stimuli occur, contribute to local calcium signaling. This may be interesting for the cell to speed up the release of vesicles, which depends on increased intracellular calcium concentration \citep{araque2014gliotransmitters}.

\begin{figure}
\includegraphics[width=0.48\textwidth]{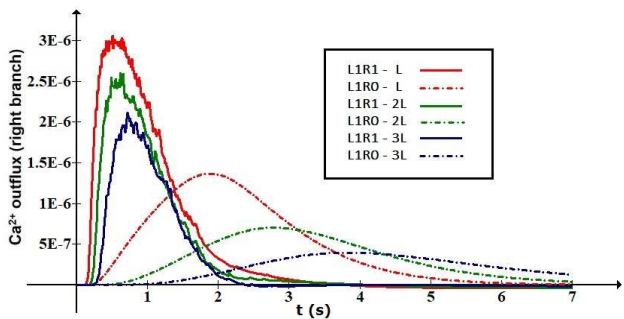}
\caption{\textbf{Instantaneous calcium flux at right branch outlet} the solid lines represent the L1R1 configurations and the dashed lines represent the L1R0 configurations. The times for the maximum points of each curve, with dimensions of the right branch of L=6\(\mu m\), 2L and 3L, are, respectively and in seconds, 0.530, 0.560 and 0.700 for configurations L1R1 and 1.890, 2.790 and 3.840 for configurations L1R0.}
\label{fig:26}       
\end{figure}

ER allocated in the branches alter the calcium signaling dynamics, as they allow the calcium concentration to increase more rapidly at the end of the distant branches from where the stimulus occurred. This is because IP\textsubscript{3} moves very quickly, reaching ERs located in distant branches long before the calcium wave produced by the ER located in the cell body reaches them. In order to better visualize this effect, a new structure was created, similar to that described in Fig. \ref{fig:18}, a central cell and two branches of length L. Simulations were performed with three sets of two configurations L1R0 and L1R1. The length of the right branch was measured L, 2L and 3L. It was possible to notice an increase in the time difference between the calcium wave at the right end, in situations in which this branch was with or without ER. The results are shown in Fig. \ref{fig:26}. Here, the solid lines represent the L1R1 configurations and the dashed lines represent the L1R0 configurations. The times for the maximum points of each curve, with dimensions of the right branch of L, 2L and 3L, are, respectively and in seconds, 0.530, 0.560 and 0.700 for configurations L1R1 and 1.890, 2.790 and 3.840 for configurations L1R0. The differences in times, between configurations L1R1 and L1R0, for dimensions L, 2L and 3L are, respectively, 1.360, 2.230 and 3.140. These points have a linear behavior and are represented by the line: \(y = (5 + 10Lx) D_{C}/D_{I}\). On the other hand, Fig. \ref{fig:23} makes it clear that, in the configuration used in this work, the final value of the accumulated calcium flow, at the exit of the right branch, does not change with the existence of ER in this branch, just contributes to accelerate calcium wave dynamics. These results suggest that we are not dealing with calcium waves, but with IP\textsubscript{3} waves with occasional geyser bursts of calcium.

\section{Conclusions}
\label{sec:15}

The present work presents three contributions to the scientific community: a mathematical model and its computational implementation, a parametric sensitivity analysis and aan analysis of the effects of ER on astrocytes branches.

Regarding the first contribution, the model was computationally constructed and implemented, to guarantee analytical flexibility of calcium signaling in astrocytes. Astrocyte geometry can be quite extensive and branched, depending only on available computational resources. In addition, the mechanisms can be easily turned on or off, allowing an analysis of their effects on specific studied phenomenon. The mathematical model also considers parameters that can be experimentally determined, including channel density and mass transfer coefficients. All these aspects are important when comparing model against experimental results; this is because there is a huge diversity of experimental results, as calcium signaling depends on location and timing of calcium fluxes \citep{volterra2014astrocyte}. This type of information is easily fed into the model, allowing a more accurate comparison.

It is important to point out that mathematical models should be made available as the scientific community improves them. \citet{manninen2017reproducibility} reported that several models are not reproducible and do not provide all the necessary information. To avoid this, the authors of the present work made available the program code so that all those who work in this area can use it, modify it, and generate the most varied results, contributing to the discussion of the role of astrocytes in brain information processing.

Regarding the parametric sensitivity analysis, more than providing a reflection about the influence of parameters and mechanisms on calcium signaling, it allowed two very pertinent conclusions. The first one is that several sets of parameters lead to the same calcium dynamics in astrocytes. This behavior is quite common in biological systems, because the system as a whole must respond in a similar way in different types of situations to which it is subjected \citep{prinz2004similar}. The second is that parameters of mathematical models cannot be regarded as absolute and require constant revision, especially when they do not come from experimental results. Several parameters used in the present work and in a sequence of literature available works were initially adjusted by \citet {keizer1992two}. Several of them were obtained from optimization of cell behavior as a whole, that is, depending on the type of experiment that was performed. In addition, the authors themselves emphasized that they chose only one of the possible sets for the assessed situation. The wide use of these parameters has led the scientific community to believe that there is no need to revise their values. According to \citet{manninen2018computational}, the models for signaling calcium in astrocytes are similar and use basically the same parameters. These values are not discussed and they are used indiscriminately. The proposition here is, given the experimental development that currently exists, specific experiments should be made for each mechanism. These new parameters can be easily aggregated into the our model, allowing for more accurate results.

Finally, the present work contributed to the analysis of ER effects on astrocytes branches. The location and function of ER in astrocytes still creates divergences of opinion in the scientific community \citep{golovina2000unloading, golovina2005visualization, holtzclaw2002astrocytes, oheim2017local}. Our results indicate that the ER has positive and negative effects when allocated in the branches, depending of the branches position in the cell. When placed next to the stimulus, they contribute to local calcium signaling; however, they affect negatively the global signaling due to the IP\textsubscript{3} degradation they promote. It is important to emphasize that the damage to global calcium signaling caused by the ER located next to the stimulus is not sufficient to prevent the signaling from advancing. By contrast, when allocated far from the stimulus site, they allow calcium signaling to quickly reach the most distant points. Thus, given brain processes speed, it is possible that astrocyte allocate ER in all branches, to accelerate the advance of calcium signaling. The consequence of the local signaling favored by the ER at branches extremities where the stimulus occurs still has to be better explored in future works.

It is important to emphasize that models for calcium signaling in astrocytes require substantial sophistication in their contribution to the understanding of astrocyte functions in cerebral information processing. Collaborative work allows us to achieve greater goals in less time. Therefore, by making our software available for the community, we welcome experimental and theoretical scientists to contribute to the development of this model and to the estimation of more appropriate parameters so all of us can make use of this versatile and useful tool.

\begin{acknowledgements}
The authors thank the Brazilian Federal Agencies, Coordination for Improvement of Higher Education Personnel, CAPES, and National Council for Scientific and Technological Development, CNPq, by scholarships for the first of them.


\end{acknowledgements}

\bibliographystyle{spbasic}      

%
%

\end{document}